# Message complexity of population protocols

**Talley Amir**
Yale, USA
https://cpsc.yale.edu/people/talley-amir
talley.amir@yale.edu

**James Aspnes**
Yale, USA
https://www.cs.yale.edu/homes/aspnes/
james.aspnes@gmail.com

**David Doty** 
University of California, Davis, USA
https://web.cs.ucdavis.edu/~doty/
doty@ucdavis.edu

**Mahsa Eftekhari**
University of California, Davis, USA
https://eftekhari.cs.ucdavis.edu/
mhseftekhari@ucdavis.edu

**Eric Severson**
University of California, Davis, USA
https://www.math.ucdavis.edu/~severson/
eseverson@ucdavis.edu

## Abstract

The standard population protocol model assumes that when two agents interact, each observes the entire state of the other agent. We initiate the study of **message complexity** for population protocols, where the state of an agent is divided into an externally-visible **message** and an internal component, where only the message can be observed by the other agent in an interaction.

We consider the case of $O(1)$ message complexity. When time is unrestricted, we obtain an exact characterization of the stably computable predicates based on the number of internal states $s(n)$: If $s(n) = o(n)$ then the protocol computes semilinear predicates (unlike the original model, which can compute non-semilinear predicates with $s(n) = O(\log n)$), and otherwise it computes a predicate decidable by a nondeterministic $O(n \log s(n))$-space-bounded Turing machine. We then introduce novel $O(\mathrm{polylog}(n))$ expected time protocols for junta/leader election and general purpose broadcast correct with high probability, and approximate and exact population size counting correct with probability 1. Finally, we show that the main constraint on the power of bounded-message-size protocols is the size of the internal states: with unbounded internal states, any computable function can be computed with probability 1 in the limit by a protocol that uses only *1-bit* messages.

**2012 ACM Subject Classification** Theory of computation → Distributed algorithms

**Keywords and phrases** Population protocols, junta election, counting, Turing machine simulation

**Funding** The second author was supported by NSF award CCF-1650596. The third, fourth, and fifth authors were supported by NSF awards 1619343, 1844976, 1900931.

# 1 Introduction

Population protocols, introduced by Angluin, Aspnes, Diamadi, Fischer, and Peralta [6], are a class of algorithms that model ad hoc networks of finite-state mobile agents. At each step, a pair of agents is picked uniformly at random to interact, each observing the other's state and updating its own state in response. The original model [6] limited the agents to $O(1)$ states, independent of the population size $n$. This limited the computational power (to only semilinear predicates [7]) and the time efficiency possible in performing fundamental tasks (such as the linear-time lower bound for leader election [29]).

Recent work has generalized to $\omega(1)$ states, increasing with $n$, finding more efficient algorithms for fundamental tasks (e.g., [1–4, 16, 32, 33, 37, 48]). Is the improved performance a consequence of higher communication throughput or higher local storage capacity?

The original model supposes that agents can view the entirety of the other's local state upon interacting with another agent, which we call an **open** protocol. We introduce a new variant of this model that draws a distinction between the state of the agent and the segment of the state that is externally visible to its interacting partner, called the **message**. This variant generalizes previous work in the context of consensus that examines the particular case of **binary signaling** [5, 42], where the message is limited to a single bit. We study the computational power of population protocols that have $O(1)$ message complexity and varying local state complexity, ranging from $O(1)$ to unbounded.

## 1.1   Motivation

The population protocol framework was conceived to model passively mobile ad hoc sensor networks. In this setting the amount of communication bandwidth can be a tighter constraint than the local computation performed by a sensor. These two constraints—bandwidth efficiency and energy efficiency—are viewed as distinct in the networking literature. In some scenarios it makes more sense to optimize for one or the other, or to strike a balance [24, 38, 49]. The restriction to $O(1)$ messages but $\omega(1)$ internal states is germane when the communication in an interaction is more costly than the accompanying local computation.

Synthetic chemistry is another domain in which population protocols are an appropriate abstract model of computation. They are a subclass of chemical reaction networks, which are known to have similar computational power [22, 45]. Using a physical primitive known as **DNA strand displacement** [51], *every* chemical reaction network with $O(1)$ species (**states** in the language of population protocols) can be theoretically implemented by a set of DNA complexes [46], justifying the use of chemical reactions as an implementable programming language. Using this approach, nontrivial chemical systems have been synthesized in the wet lab, resulting in pure DNA implementations of a chemical oscillator [47] and the "approximate majority" population protocol [9, 23]. Some theoretical [43] and experimental [50] systems are able to assemble unbounded-length heteropolymers such as DNA in an algorithmic way. For such systems, reactions may best be modeled as allowing arbitrarily many states (exponential in the polymer length) but only $O(1)$ messages modeling the smaller "locally visible region" near one or both ends of the polymer.

Finally, our model of $\omega(1)$ internal states and $O(1)$ external messages is a natural mathematical intermediate between the original $O(1)$ states/messages model and the more recent $\omega(1)$ states/messages model. Because population protocols with superconstant states and messages are provably more powerful [21], it is intrinsically interesting to determine how powerful this new intermediate model is.

## 1.2   Our Contribution

We introduce this new variant of population protocols and show three main results:

We first (Section 3) completely resolve the question of the computational power of $O(1)$ messages, with Theorem 3.2. In the positive direction, with poly($n$) states ($O(\log n)$ bits), we give a simulation of $\Omega(1)$-bit messages (Theorems 3.3 and 3.5). Corollary 3.9 is an asymptotically sharp negative result: $O(1)$-message, $o(n)$-state ($\log n - \omega(1)$ bits) protocols compute only semilinear predicates.

| **Problem solved** | **Pr[error]** | **Time** | **States** | $|M|$ | **Leader** |
|---|---|---|---|---|---|
| Simulate $s(n)$-state open protocol (Corollary 3.4) | 0 | $O(t_P n^2 \log s(n))$ | $O(s(n)^2)$ | $O(1)$ | Yes |
| Junta election (Theorem 5.1) | $> 0$ | $O(\log^2 n)$ | $O(\log^2 n)$ | 1-bit | No |
| Compute $n$ (Theorem 5.7) | $> 0$ | $O(\log^2 n)$ | $O(n \log^2 n)$ | $O(1)$ | Yes |
| Compute $\log n$ (Corollary 5.8) | $> 0$ | $O(\log^2 n)$ | $O(\log n)$ | $O(1)$ | Yes |
| Stably compute $n$ (Corollary 5.9) | 0 | $O(\log^2 n)$ | $O(n^4 \log^4 n)$ | $O(1)$ | Yes |
| Leaderlessly compute $n$ (Corollary 5.10) | $> 0$ | $O(\log^2 n)$ | $O(n \text{ polylog}(n))$ | $O(1)$ | No |
| Leaderlessly compute $\log n$ (Corollary 5.10) | $> 0$ | $O(\log^2 n)$ | $O(\text{polylog}(n))$ | $O(1)$ | No |
| Compute $d$-input predicate (Corollary 5.11) | $> 0$ | $O(d \log^2 n)$ | $O(n^d \log^2 n)$ | $O(1)$ | Yes |
| TM simulation (Theorem 6.5) | 0 | unbounded | unbounded | 1-bit | No |

**Table 1** Summary of positive results: Above, the event of "not error" means that the answer is correct *and* the stated time and state bounds hold, unless the error probability is 0, in which case it refers only to the output being correct. In that case, the time and state bounds are in expectation, but still hold with high probability: in all cases, the probability of error is $O(1/n)$. (It should be straightforward to extend to $O(1/n^k)$ for any $k$, but for simplicity in proof statements we fix $k = 1$. We rely on [15, Theorem 1] that, as stated, holds only for a fixed exponent $k$, though it seems a more detailed analysis could achieve arbitrary $k$.) Note that when the probability of computing the correct output is 1 (i.e. the protocol stabilizes), the Time column denotes time to convergence. State complexities are accurate with high probability. $|M|$ is the number of messages, either a constant larger than 1, or exactly 2 (1-bit). Compute $\log n$ means computing either $\lfloor \log n \rfloor$ or $\lceil \log n \rceil$. In the first row $t_P$ is the expected convergence time for $P$.

Secondly (Section 5), we focus on time-efficient computation. We develop novel $O(\log^2 n)$-time algorithms for junta election (the key primitive to leader election) and exact population size counting (naturally suited to this model, where $O(n)$ local states and $O(1)$ messages are the minimal power to make this problem solvable). The counting protocol can specialize with fewer states to estimate the size (count $\log n$), and also generalize with more states to count the entire input configuration (so any predicate can be locally computed).

Thirdly (Section 6), we explore the extreme limits of the model where message complexity is limited to 1 bit. We construct a 1-bit broadcast primitive, showing it is powerful enough to simulate a Turing Machine with probability 1 correctness using unbounded local memory.

## 1.3 Comparison to existing work and new techniques required

Most protocols using $\omega(1)$ states [1–3, 13, 14, 17–20, 26, 32, 33, 39, 41, 48] crucially use $\omega(1)$-size messages. Key transitions in such protocols involve comparing two integers/ids of size $\omega(1)$ in a single step, which is not possible with $O(1)$-size messages. Sending a superconstant-size message over multiple interactions is not efficient (though it is a trick we employ for unbounded time results such as Theorem 3.3), since there is not enough time for the two agents to wait for another interaction (which takes $\Theta(n)$ expected time), nor is there any way to distinguish each other in future interactions. We introduce new techniques that rely on timing of internal counters to get around this limitation.

Protocol 2, "junta election", is our primary fast leaderless protocol, used to make other leader-driven protocols leaderless. It elects a **junta**, a group of $O(\sqrt{n})$ agents, in $O(\log^2 n)$ time. As with many other existing protocols [16, 32, 33, 48], this is used to drive a junta-driven **phase clock** [8] that allows agents to synchronize in a downstream computation. The cited protocols have agents choose an integer "level" $\ell$, propagating by epidemic the maximum level ($\Theta(\log \log n)$ [16, 32, 33] as in our case, or $\Theta(\log n)$ [48]). Agents who chose the maximum level are in the junta. Lacking the ability to communicate the levels in 1-bit messages, we rely on **timing** of internal counters of agents to detect whether a higher level exists: Agents with level $\ell$ count up to $\approx 4^\ell$, (roughly) telling all other agents to continue counting up, and stop at $\approx 4^\ell$, unless another agent (with high probability with a higher level) tells them to continue counting. The actual details are more involved and require intricate choice of timing and analysis to conclude that all agents with high probability stop at the same counter value.

We push the technique of communication via timing further, showing that only *1-bit* messages suffice to elect a leader, broadcast arbitrary messages, and simulate a Turing Machine.

The **majority** problem means deciding which of two opinions in a population is more numerous. Existing protocols [2, 4, 13, 14, 19] use an $O(1)$-message "doubling/cancelling" technique, which works on top of a synchronization primitive, but these protocols use $\omega(1)$ messages to achieve synchronization. Thus, our $O(1)$-message junta-election protocol can be composed with the doubling/cancelling technique to give a high-probability, *uniform* (unlike [2,4,13,19]) $O(1)$-message majority protocol. In fact, the doubling/cancelling technique can be viewed as another special case of our general population counting algorithm.

All of our protocols are **uniform** (requiring no estimate of $n$), in contrast to several existing $\omega(1)$-state protocols [1–4, 13, 17, 19, 20, 33, 39, 41, 48]. Many of our protocols could be simplified greatly by allowing nonuniformity. Briefly, an estimate of $\log n$ within a constant factor allows agents to run a **leaderless phase clock** in which they count up to $c \cdot \log n$ (for some large constant $c$), which aids in synchronizing agents in rounds $r = 0, 1, 2, \ldots$ based on the number of times they have counted from 0 up to $c \log n$; the lack of such synchronization is a major challenge in devising correct, efficient $O(1)$-message protocols.

## 2   Model

We write $\log n$ to denote $\log_2 n$, and $\ln n$ to denote $\log_e n$. The original population protocol model [6] involves a population of $n$ **agents**, each of which holds a **state** in a **state space** $Q$. Interactions between agents update the states of both agents according to a **transition function** $\delta$ that takes both states as input and returns new states for both agents as output. Interactions are asymmetric: in each interaction, one of the agents is the **initiator** of the interaction, and one the **responder**.

**Uniform versus nonuniform protocols.**

The original model [6] considered $|Q|$ to be constant, but later works [1–4, 13, 17, 19, 27, 32, 39, 41, 48] let $|Q|$ scale with the population size $n$. Some protocols have been **nonuniform**, where $\delta$ implicitly depends on $n$, but it is preferable for a protocol to be **uniform**, with an identical $\delta$ over all population sizes $n$. The work of [21] gave one formalization of uniform population protocols with superconstant state space.[1]

---

[1] Essentially, they model agents as Turing Machines, which are able to exchange messages from special message tapes during an interaction. A protocol consists of the Turing Machine transition rules, that are executed by all agents in all population sizes. The space bound is the maximal space used by any Turing Machine in any execution. Thus $\delta$ is being effectively computed by some space-bounded Turing

We consider a refinement of the model in which the state of an agent is explicitly divided into an internal component that is not visible to other agents, and an external component that is. The internal component of the state is drawn from the state space $I$ and the external component, or **message**, is drawn from a message space $M$. The set of states $Q$ is the Cartesian product $I \times M$. The transition function $\delta$ is modified to enforce the restriction that an agent in an interaction cannot observe the internal state of the other agent: $\delta$ is now a function from $Q \times M \times \{\text{initiator}, \text{responder}\}$ to $Q$. When an agent in state $q_1 = \langle i_1, m_1 \rangle$ initiates an interaction with an agent in state $q_2 = \langle i_2, m_2 \rangle$, the new states of the agents are given by $q_1' = \langle i_1', m_1' \rangle = \delta(q_1, m_2, \text{initiator})$ and $q_2' = \langle i_2', m_2' \rangle = \delta(q_2, m_1, \text{responder})$.

The set of producible states $Q(n)$ and the set of producible messages $M(n)$ can both depend on $n$. The function $s : \mathbb{N} \to \mathbb{N}$ defined as $s(n) = |Q(n)|$ is the **state complexity** [2] of a population protocol. The function $n \mapsto |M(n)|$ is the **message complexity**. If $|I| = 1$ and each agent's state is merely defined by its message (the original model [6] and its superconstant state generalization), we say the protocol is **open**, so $|Q(n)| = |M(n)|$ for all $n$. We will mostly be interested in population protocols with modest state complexity (at most polynomial in $n$, and often only polylogarithmic in $n$) and constant message complexity. Given two functions $s, m : \mathbb{N} \to \mathbb{N}$, a $s(n)$**-state,** $m(n)$**-message population protocol** is one with state complexity $s$ and message complexity $m$. Note that the complexity bounds we discuss are *worst-case*: $s(n)$ is the most number of states that can be produced in any population of size $n$ under any execution.

We will also place high probability bounds on the state complexity (such as Protocol 2 where each agent generates a geometric random variable, which may take on any positive integer value). These are not statements about the set of producible states, so our impossibility results (Theorem 3.8) on state and message complexity do not apply.

*Problems solved by population protocols.* A **configuration** gives the state of all agents. Population protocols have some problem-dependent notion of "correct" configurations. For example, for **leader election** a configuration with a single leader is correct. For computation of a **predicate** $\phi : \mathbb{N}^d \to \{\text{yes}, \text{no}\}$ (a.k.a., **decision problem**), the initial state of each agent is from a $d$-element subset $\Sigma$ of states, states are partitioned into two subsets representing "yes" and "no", and a configuration is correct if all agents give the answer $\phi(\vec{i})$, where $\vec{i} \in \mathbb{N}^d$ represents the initial counts of agents in each state in $\Sigma$. A population protocol is **leader-driven** if its states have a Boolean field $\text{leader} \in \{L, F\}$ (i.e. the state set $Q = \{L, F\} \times Q'$), such that in every valid initial configuration, exactly one agent has $\text{leader} = L$.

*Time complexity.* For measuring time complexity, we assume **random scheduling**, where at each interaction two agents are chosen uniformly at random from all $n(n-1)$ possible ordered pairs of agents. Time complexity is defined by **parallel time**, the number of interactions divided by $n/2$ which we henceforth simply refer to as **time**[3].

---

Machine, and their model counts the space cost of this local computation. Formally, our model can be viewed as a special case of this model, where we restrict the message tape to $O(1)$ bits. For the purpose of brevity, however, we give a more intuitive definition based on a transition function $\delta$. Also, to talk about time complexity, we use the standard uniform random interaction scheduler and assume the local computation happens instantaneously.

[2] This definition of state complexity abstracts away the space used for the local computation of $\delta$ as counted in [21]. It can be thought of as a simpler information-theoretic measure of how many different memory configurations agents can be in before and after—but not during—their transitions. Also, the space overhead to compute $\delta$ will always use $O(\log |Q|)$ bits, so the asymptotic size $\Theta(\log |Q|)$ of the state space in bits will be unchanged. Because the results of [21] are all asymptotic statements about the number of bits of memory, they can apply directly.

[3] This scaling $n/2$ has each agent participate in an expected $k$ interactions after $k$ time steps, convenient for later proofs.

*Convergence/stabilization.* A configuration $\vec{c}$ is **stably correct** if every configuration reachable from $\vec{c}$ is correct. An execution $\mathcal{E} = (\vec{c}_0, \vec{c}_1, \ldots)$ is picked at random according to the scheduler explained above. We say $\mathcal{E}$ **converges** (respectively, **stabilizes**) at interaction $i \in \mathbb{N}$ if $\vec{c}_{i-1}$ is not correct (resp., stably correct) and for all $j \geq i$, $\vec{c}_j$ is correct (resp., stably correct). The **(parallel) convergence/stabilization time** of a protocol is the number of iterations to converge/stabilize, divided by $n/2$. Convergence can happen strictly before stabilization, although a protocol with finite reachability (i.e., for each $\vec{c}$, finitely many configurations are reachable from $\vec{c}$) converges from $\vec{c}$ with probability $p \in [0, 1]$ if and only if it stabilizes from $\vec{c}$ with probability $p$. For a computational task $T$ equipped with some definition of "correct", we say that a protocol **stably computes** $T$ **with probability** $p$ if, with probability $p$, it stabilizes (equivalently, converges). Note that detecting that the protocol has reached a correct output configuration is generally not possible.

**Other notions of computation**

The problems we consider in this paper (predicate computation, junta and leader election, and Turing machine simulation) all consider the setting of a constant-sized input alphabet $\Sigma$, where each agent starts with a symbol from $\Sigma$. Since we allow $\omega(1)$ internal states, this could in principle model problems allowing non-constant input per agent, e.g., agents each start with an integer, and we want to calculate their median. There is also a notion of **function** computation with population protocols [12, 22, 28] in which the natural number output, rather than being fully written into the internal state of each agent, is distributed across the population "in unary", i.e., the output is $k$ if exactly $k$ agents are in a special state state $Y$. However, we do not consider such problems in this paper.

## 3 Computability with unrestricted time

In this section we study $s(n)$-state, $O(1)$-message protocols, when the time is not restricted. Theorem 3.2 is our main result in this section, which completely characterizes the power of such protocols in terms of the number of bits required to store the states.

Let $\mathsf{CMPP}(f(n))$ be the set of all predicates stably computed by an $s(n)$-state, $O(1)$-message population protocol, where $s(n) = 2^{O(f(n))}$ (using $O(f(n))$ bits of memory) [4]. Let $\mathsf{SNSPACE}(g(n))$ be the set of all predicates $\phi : \mathbb{N}^d \to \{0, 1\}$ decidable by a nondeterministic $O(g(n))$-space-bounded Turing machine, when inputs are given in unary.[5] The results of [21] considered a similar complexity class $\mathsf{PMSPACE}(f(n))$ of stably computable predicates using $O(f(n))$ bits of memory and $O(f(n))$ bit messages.[6] Let $\mathsf{SL}$ be the set of all **semilinear** predicates [7]. Their main result is the following characterization:

▶ **Theorem 3.1** ( [21]). *Let $f : \mathbb{N} \to \mathbb{N}$. If $f(n) = o(\log \log n)$, then $\mathsf{PMSPACE}(f(n)) = \mathsf{SL}$. If $f(n) = \Omega(\log n)$, then $\mathsf{PMSPACE}(f(n)) = \mathsf{SNSPACE}(n \cdot f(n))$.*

---

[4] Our model formally requires the local computation of $\delta$ to take $O(f(n))$ bits, so here $f(n)$ is the full memory bound on local computation, as in [21].
[5] In [21] these are called **symmetric** predicates on the assumption that the $d$ counts in $\vec{i} \in \mathbb{N}^d$ are presented to the Turing machine as a $\|\vec{i}\|$-length string of symbols from an input alphabet $\Sigma$ with $|\Sigma| = d$, with the same answer on all permutations of the string.
[6] In fact, to obtain their positive result for large space bounds, they do not need fully open protocols. Their simulation of nondeterministic $nf(n)$-space-bounded Turing machines just requires $O(\log n)$ bit messages to exchange unique IDs, even if $f(n) = \omega(\log n)$.

Since the memory is expressed in Theorem 3.1 as number of **bits** (exponentially smaller than number of **states**), the multiplicative constants hidden in the $O()$ notation become polynomial-factor terms in number of states. Theorem 3.2 is a similar dichotomy theorem for $O(1)$-message population protocols, which is sharper in that it holds for **all** values of $f(n)$.

▶ **Theorem 3.2.** *Let* $f : \mathbb{N} \to \mathbb{N}$. *If* $f(n) = o(\log n)$, *then* $\mathsf{CMPP}(f(n)) = \mathsf{SL}$, *otherwise* $\mathsf{CMPP}(f(n)) = \mathsf{SNSPACE}(n \cdot f(n))$.

**Proof.** First note that $2^{O(f(n))}$-state $O(1)$-message population protocols are a special case of the Passively Mobile Machines from [21] with space bound $f(n)$ (since we assume the space overhead to compute $\delta$ is $O(f(n))$ bits[7]). Thus $\mathsf{CMPP}(f(n)) \subseteq \mathsf{PMSPACE}(f(n))$.

When $f(n) = \Omega(\log n)$, we will show via Theorem 3.5 and Theorem 3.3, that $2^{O(f(n))}$-state $O(1)$-message population protocols can simulate a open protocols, with a polynomial state overhead (ie. a constant overhead in $f(n)$ which does not change the definition of the complexity classes). The ability to simulate large messages then implies $\mathsf{PMSPACE}(f(n)) \subseteq \mathsf{CMPP}(f(n))$, and then using Theorem 3.1 we have $\mathsf{CMPP}(f(n)) = \mathsf{PMSPACE}(f(n)) = \mathsf{SNSPACE}(n \cdot f(n))$.

Finally, when $f(n) = o(\log n)$, we have $s(n) = 2^{O(f(n))} = o(n)$ and we will show via Corollary 3.9 that $\mathsf{CMPP}(f(n)) = \mathsf{SL}$. Note that our necessary condition $s(n) = o(n)$ in Corollary 3.9 is actually even sharper than $\log(s(n)) = o(\log n)$. ◀

## 3.1 Leader-driven $O(s(n)^2)$-state, $O(1)$-message protocols can simulate open $s(n)$-state protocols

In this section we show that $O(s(n)^2)$-state, $O(1)$-message, leader-driven protocols can simulate $s(n)$-state open protocols (whether leader-driven or not). Thus, allowing a leader and ignoring quadratic differences in state complexity (see discussion of quadratic blowup below ), there is no difference whatsoever between the computational power of $O(1)$-message protocols and open protocols. Theorem 3.3 proves the general case of $m(n)$-message protocols, and Corollary 3.4 is the special case of **open** protocols, where $s(n) = m(n)$. The simulation incurs a time slowdown of factor $n^2 \log m(n)$, where $n$ is the population size and $m(n) \leq s(n)$ is the message complexity of the simulated protocol, so it helps port computability results from the open protocol model, but not sublinear time results.

**Quadratic state blowup in Theorem 3.3.**

The quadratic state blowup of Theorem 3.3 is an artifact of definitional choice, in a sense, owing to each agent $a$ needing to write down the state of another agent $b$, bit by bit over many interactions, before $a$ can execute the transition $\delta$. However, the model from [21] explicitly counts the space required to store the other agent's message against the total space required, so there is no space blowup in that case.

Intuitively, the construction of Theorem 3.3 chooses two agents to "mark" as initiator and responder, which then successively pass a bit string as they interact, until they have transmitted the full message of size $\log m(n)$ bits. Crucially, starting with a leader allows only one simulated transition to be taking place at a time.

---

[7] Note even if our definitions were more powerful and the space overhead to compute $\delta$ was as large as $O(nf(n))$ bits, we could still make the argument of Theorem 5 of [21] to conclude an $O(nf(n))$ nondeterministic Turing Machine can simulate an $2^{O(f(n))}$-state $O(1)$-message population protocols, and Theorem 3.2 would still hold.

**Simulation of a population protocol by another.**

Formal definitions of simulation in population protocols exist [36, 44] (for the strictly more general model of chemical reaction networks), but such definitions are complex and have to cover many corner cases when applied to arbitrary systems. Since we study just a single simulation construction in Theorem 3.3, we avoid a completely formal definition in this paper. Let $P, S$ be population protocols and $\vec{c}_P, \vec{c}_S$ be configurations of $P$ and $S$, respectively. Intuitively, we say that $S$ from $\vec{c}_S$ **simulates** $P$ from $\vec{c}_P$ if, for every execution $\mathcal{E}_P$ of $P$ starting at $\vec{c}_P$, there is an execution $\mathcal{E}_S$ of $S$ starting at $\vec{c}_S$ that "looks like" $\mathcal{E}_P$, and furthermore every **fair** execution $\mathcal{E}_S$ of $S$ starting at $\vec{c}_S$ "looks like" some fair execution $\mathcal{E}_P$ of $P$ starting at $\vec{c}_P$.

Here, "looks like" is a tricky concept that can be formalized in a few ways. Intuitively, we imagine that the states of $P$ are projections of the states of $S$, i.e., each state of $S$ is a pair $(p, e)$, where $p$ is a state of $P$ and $e$ is extra "overhead" information that $S$ requires for the simulation. Furthermore, if we project states from $\mathcal{E}_S$ onto only the first state element $p$ for each agent, and we remove those transitions that appear null from the point of view of $P$ (i.e., the $p$ portion of the state does not change in any agent), and we similarly remove null transitions from $\mathcal{E}_P$, then the resulting executions $\mathcal{E}'_S$ and $\mathcal{E}'_P$ are identical (i.e., go through the exact same sequence of configurations of $P$).

▶ **Theorem 3.3.** *For every $s(n)$-state, $m(n)$-message protocol $P$, there is a leader-driven, $O(s(n) \cdot m(n))$-state, $O(1)$-message[8] protocol $S$ that simulates $P$, and each interaction of $P$ takes expected $O(n^2 \log m(n))$ interactions of $S$ to simulate.*

**Proof.** Let $M_P$ be the messages of the simulated protocol $P$, and $\delta_P : Q_P \times M_P \times \{\mathsf{initiator}, \mathsf{responder}\} \to Q_P$ be its transition function. Intuitively, we will simulate $\delta_P$ by marking two agents to exchange bit strings over $O(\log |Q_P|)$ interactions, so each learns the message of the other and can locally compute $\delta_P$.

We now define $Q = I \times M$, the state set of the simulating protocol $S$. The internal state $I$ contains two fields:

1. a value $p \in Q_P$ representing the state of this agent in the simulated protocol $P$
2. a value $m_o \in M_P$ representing a message of the "other" agent. It is easiest to think of the messages in $M_P$ as binary strings in $\{0, 1\}^*$, because this field will be built up bit-by-bit in interactions to learn the other agent's full message. Thus, $\lambda$ (the empty string) will represent having no information about any other agent's message.

$S$ is leader-driven, so there is a field $\mathsf{leader} \in \{L, F\}$ within the message state $M$. $M$ also contains a field $\mathsf{token} \in \{\mathit{True}, \mathit{False}\}$, a field $\mathsf{mark} \in \{r, i, u\}$ (responder,initiator,unmarked), and a field $\mathsf{bit} \in \{0, 1, \mathtt{end}\}$.

To represent an initial configuration $\vec{c}_P$ of $P$, we define the initial configuration $\vec{c}_S$ (with $\|\vec{c}_S\| = \|\vec{c}_P\|$) of $S$ as follows. Each agent in $S$ has its field $p$ representing a state in $Q_P$ in the obvious way, $\lambda$ for $p_o$, $\mathsf{token} = \mathit{False}$ and $\mathsf{mark} = u$ (unmarked). Because $S$ is leader-driven, exactly one agent starts with $\mathsf{leader} = L$.

We now describe the transition function $\delta$ of $S$, at a high level. All non-null interactions are between an agent with $\mathsf{token} = \mathit{True}$. The leader $L$, on its first interaction, immediately becomes a follower $F$, and the other agent sets $\mathsf{token} = \mathit{True}$.

---

[8] The message bound is an absolute constant that does not depend on $P$. By inspecting the messages as defined in the proof, it is at most $2 \cdot 2 \cdot 3 \cdot 3 = 36$ total messages, though some combinations of fields never appear together, so can be reduced somewhat.

If the initiator agent has token = *True* with mark $u$, then it marks itself as $i$ and the other agent with $r$; otherwise it marks itself as $r$ and the other agent with $i$. The other agent now knows it is a receiver/initiator in the simulated transition. All agents have null transitions now, except for two marked agents. (They could be picked in the opposite order on subsequent transitions and still carry out the following protocol; the initiator-responder distinction in $S$ only matters for the very first transition of $S$ simulating a transition of $P$.)

Now, the responder and initiator communicate their messages from $M_P$ one bit at a time, storing the other agent's message by appending the received bits to the field $p_o$, using the field bit, sending the value end to indicate their message string has ended. Once both agents have received the other's full message, they can compute $\delta_P$ to change their simulated state $p$. Finally, they both set their mark value to $u$, and the agent with token = *True* sets token = *False* and leader = $L$. It is now ready to pass the token to the next agent it sees to simulate another transition of $P$.

Note that there is one form of asymmetry in the sense that no agent can have the token twice in a row; hence the probabilities of transitions $S$ simulates are different from the original transition probabilities in $P$. Still, at each new step in the simulation (when an agent who had the token sets leader = $L$ and then passes off the token), every possible transition can be simulated (since the new token recipient can pick the old token holder to mark for the next interaction as well, giving them either $r$ or $i$). After this nondeterministic choice, the protocol $S$ stably simulates the transition is has committed to by the assignment of mark.

Since all possible transitions can be chosen at each step, and the transition will be stably executed (in expected $O(n^2 \log m(n))$ interactions for the $i$ and $r$ mark to meet enough to pass the whole message), $S$ faithfully simulates $P$. ◀

We note that execution probabilities are not preserved by this simulation. The agent with the token in the current simulated interaction is half as likely to be chosen in the next simulated interaction as the rest of the population. Section 3 focuses on probability-1 results, which are robust to this change. By passing the initial token a larger number of times, the agent-pair probabilities are closer, but not equal, to uniform. We leave open whether there is a simulation as in Theorem 3.3 that exactly preserves execution probabilities.

The next corollary applies to **open** protocols, where each agent's message is its full state.

▶ **Corollary 3.4.** *For every $s(n)$-state, open population protocol $P$, there is a leader-driven, $O(s(n)^2)$-state, $O(1)$-message population protocol $S$ that simulates $P$, and each interaction of $P$ takes expected $O(n^2 \log s(n))$ interactions of $S$ to simulate.*

It is known that $\Omega(\log n)$-state open protocols have computational power beyond that of $O(1)$-state protocols (limited to semilinear predicates [7] and functions [22]), and Corollary 3.4 grants this same computational power to leader-driven $O(1)$-message protocols. Theorem 3.8 in subsection 3.4 shows that Corollary 3.4 crucially depends on the assumption of an initial leader in the simulating protocol, by demonstrating that **leaderless** $O(1)$-message, $o(n)$-state protocols are no more powerful than $O(1)$-state open protocols.

## 3.2 Leader election can be composed with leader-driven, $s(n)$-state, $O(1)$-message protocols using $O(n^3 \log n)$ state overhead

Leader election is possible in linear time with 1-bit messages by "fratricide": $\ell, \ell \to \ell, f$. A downstream leader-driven protocol $P$ will not work unaltered if composed with this leader election, because the presence of multiple leaders prior to convergence causes incorrect transitions of $P$. A straightforward fix using $O(n)$ messages involves exact size counting via transitions $\ell_i, \ell_j \to \ell_{i+j}, f_{i+j}$ (requiring $\Omega(n)$ messages) and each transition between agents with respective values $n$ and $i < n$ resets the latter agent to its initial state in $P$. At the moment the last agent is reset with value $n$, the protocol at that point faithfully executes a **tail** of an execution of $P$ from $\vec{i}$, i.e., an execution starting at a configuration $\vec{c}$ reachable from $\vec{i}$. Thus if $P$ is correct with probability 1, the composed protocol is also correct with probability 1. Theorem 3.5 shows how to achieve a similar "composition by resetting" strategy using only $O(1)$ messages.

■ **Protocol 1** *StablyComposableLeaderElection*(Agent $v$ seeing message $m$). $P$ is the downstream protocol with state set $Q_P$, message set $M_P$, and transition function $\delta_P : Q_P \times M_P \times \{\text{initiator}, \text{responder}\} \to Q_P$. $P$ is **leader-driven**, with a field $\text{leader}_P \in \{L, F\}$ and possible input states $\Sigma_P$. State set $Q$ of composed protocol $S$ is $Q_S \times Q_P$, where $Q_S = \mathbb{N} \times \mathbb{N} \times \mathbb{N} \times \Sigma_P \times M_S$ is the **overhead** of the composition. The fields are named $c_0 \in \mathbb{N}$, $c_1 \in \mathbb{N}$, $\text{count} \in \mathbb{N}$, $i_P \in \Sigma_P$ (representing the input symbol for protocol $P$), and $M_S$ has fields $\text{role} \in \{\ell_S, f_S\}$, $\text{phase} \in \{0, 1\}$ and $\text{signal} \in \{restart, go\}$.

1 **initial state of agent** $v$: $c_0 = c_1 = \text{count} = 0$, $i_P = q_P = $ **initial input state of** $v$ **in** $P$, $\text{role} = \ell_S$, $\text{phase} = 0$, $\text{signal} = \boldsymbol{restart}$
2 **if** $v.\text{signal} = go$ **and** $m.\text{signal} = go$ **then**
3   $\quad v.q_P \leftarrow \delta_P(v.q_P, v.m_P, \text{initiator if } v \text{ is initiator, else responder})$
4 **if** $v.\text{role} = \ell_S$ **and** $m.\text{role} = \ell_S$ **then**
5   $\quad$ **if** $v$ *is* responder **then**
6   $\quad\quad v.\text{role} \leftarrow f_S$ ;                     // fratricide leader election
7   $\quad$ **else**
8   $\quad\quad$ reinitialize $v$ ;          // surviving leader goes back to initial state
9 **else if** $v.\text{role} = \ell_S$ **and** $m.\text{role} = f_S$ ;    // base-station counting from [10]
10 **then**
11   $\quad$ **if** $b = v.\text{phase} = m.\text{phase}$ **then**
12   $\quad\quad v.\text{count} \leftarrow 0$
13   $\quad\quad v.c_{1-b} \leftarrow v.c_{1-b} + 1$
14   $\quad\quad$ **if** $v.c_b = 0$ **then**
15   $\quad\quad\quad v.\text{signal} \leftarrow restart$ ;   // population estimate $c_0 + c_1$ has increased
16   $\quad\quad\quad v.q_P \leftarrow v.i_P$
17   $\quad\quad\quad v.\text{leader}_P \leftarrow L$
18   $\quad\quad$ **else if** $v.c_b > 0$ **then**
19   $\quad\quad\quad v.c_b \leftarrow v.c_b - 1$
20   $\quad\quad\quad$ **if** $v.c_b = 0$ **then**
21   $\quad\quad\quad\quad v.\text{signal} \leftarrow go$ ;       // all counted agents have been restarted
22   $\quad$ **else if** $v.\text{count} \geq 6 c_{1-b} \ln c_{1-b} + 1$ **then**
23   $\quad\quad v.\text{count} \leftarrow 0$
24   $\quad\quad v.\text{phase} \leftarrow 1 - v.\text{phase}$
25   $\quad$ **else if** $c_{v.\text{phase}} = 0$ **then**
26   $\quad\quad v.\text{count} \leftarrow v.\text{count} + 1$
27 **else if** $v.\text{role} = f_S$ **and** $m.\text{role} = \ell_S$ **then**
28   $\quad v.\text{phase} \leftarrow 1 - m.\text{phase}$
29   $\quad v.\text{signal} \leftarrow m.\text{signal}$
30   $\quad$ **if** $v.\text{signal} = restart$ **then**
31   $\quad\quad v.q_P \leftarrow v.i_P$
32   $\quad\quad v.\text{leader}_P \leftarrow F$

▶ **Theorem 3.5.** *For any leader-driven, $s(n)$-state, $O(1)$-message protocol $P$, there is a leaderless, $O(s(n)n^3 \log n)$-state, $O(1)$-message protocol $S$ (Protocol 1) that, after $O(n \log n)$ expected time, executes a tail of an execution of $P$.*

**Proof sketch.** Briefly, we elect a leader in $O(n)$ time by fratricide. The leader counts the followers using the $O(n \log n)$-time counting protocol of [10], which is self-stabilizing under the assumption (proven necessary in [11]) of a "base-station": a leader that is initialized (starts in a pre-determined state), though other agents can start in arbitrary states. That paper does not use the term "self-stabilizing" explicitly, but the assumptions we claimed are stated in their introduction. In our case, the remaining leader resets to the initial state whenever it kills another leader, so after the last such transition, the unique leader is initialized, and other agents are in arbitrary states, exactly the setting handled in [10]. Whenever the leader's population count increases, it tries to reset each follower. Once the leader has counted the whole population and knows $n$, its count will never change, and it will reset every follower to its initial state of $P$ by direct interaction (using its knowledge of $n$ to ensure all $n-1$ followers are reset), at which point a tail of $P$ executes.  ◀

**Proof.** First observe that the field role updates via the standard "fratricide" leader election. Thus there exists some first time $t_1$, (with $E(t_1) = O(n)$), where there is a unique agent $a.\mathsf{role} = \ell_S$. By line 8, $a$ has reinitialized with $c_0 = c_1 = \mathsf{count} = \mathsf{phase} = 0$.

Now agent $a$ acts as the base-station in the self-stabilizing counting protocol of [10], communicating with the other agents via the field phase. In each phase $b \in \{0, 1\}$, $a$ counts the other agents it sees in $\mathsf{phase} = b$, moving them into phase $1-b$, decrementing its counter $c_b$ (if possible) and incrementing $c_{1-b}$. By the results of [10], the count $c_0 + c_1$ increases monotonically, and stabilizes at a maximum value of $n-1$ in $O(n \log n)$ expected time.

Let $t_2$ be the first time $c_0 + c_1 = n - 1$, with $a.\mathsf{phase} = b$. Then $c_{1-b}$ just incremented to $n-1$, and $c_b = 0$ and failed to decrement. The if condition in line 14 was true, so $a.\mathsf{signal} = \mathit{restart}$. In all future interactions, $c_0$ and $c_1$ accurately count the number of follower agents in each phase, so the condition of line 14 will never be met again.

Now consider the next time $t_3$ when $a$ changes to phase $1-b$ and brings the count $c_{1-b} = 0$. By [10] this will also take an expected $O(n \log n)$ time. Then, since time $t_2$, $a$ has interacted with all agents, who now have $v.\mathsf{signal} = \mathit{restart}$. By line 21, $a.\mathsf{signal} = \mathit{go}$ for the first time since $t_2$. Then every agent $v$ who interacts with $a$ will have $v.\mathsf{signal} = \mathit{go}$ for all future interactions.

Let $\vec{i}$ denote the configuration in protocol $P$ when every agent has their original input state $i_P$, alongside $a.\mathsf{leader}_P = L$ and $v.\mathsf{leader}_P = F$ for all $v \neq a$. Now observe that when each agent sets $v.\mathsf{signal} \leftarrow \mathit{go}$ for the last time, they have the same configuration as in $\vec{i}$. They only execute transitions in $P$ via line 3 with other agents with $\mathsf{signal} = \mathit{go}$.

Let $t_4$ be the next time when every agent has $\mathsf{signal} = \mathit{go}$, and $\vec{c}$ the configuration within $P$ at $t_4$. Then the only transitions that have made $\vec{c}$ different from $\vec{i}$ were between two agents with $\mathsf{signal} = \mathit{go}$, who began from an initialized state. Thus $\vec{c}$ is reachable from the correctly initialized configuration $\vec{i}$. Finally, all future transitions execute $\delta_P$ on both agents, so the composed protocol now exactly implements $P$, i.e., executes an execution of $P$ from $\vec{c}$, i.e., a tail of an execution from $\vec{i}$.

Note that we assume that (as is the case in most population protocols, even with $\omega(1)$ states), that only a $O(1)$-size subset $\Sigma$ of states appear in valid initial configurations, thus $i_P$ in Protocol 1 gives at most a constant-factor overhead to the simulation. If instead agents could start with more states, then the factor would be $|\Sigma_P|$.  ◀

Theorem 3.5 depends crucially on using $\geq n$ states, since Theorem 3.8 shows leaderless, $O(1)$-message, $o(n)$-state protocols are no more powerful than $O(1)$-state open protocols.

## 3.3  Deterministic Broadcast

The construction used in Protocol 1 can be modified to also give the leader the ability to stably broadcast a message to the entire population. After the last restart, the leader agent $a$ counts the entire population by moving them between phases. We can view these phases now as deterministically synchronized rounds (each expected time $O(n \log n)$ [10]). Add a field bit $\in \{0, 1\}$ to the message. The leader $a$ can then communicate a bit string to the population by sending one bit during each round. This lets the entire population stably compute the population size $n$, by having the leader send $n$ as a bit string in $O(\log n)$ rounds (stabilizing in expected $O(n \log^2 n)$ time). It uses $O(\log n)$ state overhead to store the number of bits it has broadcast, so $O(n^3 \log^2 n)$ states total. We can thus conclude:

▶ **Corollary 3.6.** *There is an $O(n^3 \log^2 n)$-state, $O(1)$-message protocol that stably computes the population size $n$ (storing in every agents state), in expected $O(n \log^2 n)$ time.*

Building on the ideas used in Corollary 3.6, we can have the leader assign unique IDs to the agents, for example marking a new unmarked agent in each synchronized round. On top of this deterministic broadcast primitive, we could set up a nondeterministic Turing Machine simulation equivalent to the construction in [21]. This gives a direct constructive proof of Theorem 3.2, rather than relying on the simulation arguments via Corollary 3.4 and Theorem 3.5.

## 3.4  Leaderless $o(n)$-state, $O(1)$-message protocols compute only semilinear predicates

Theorem 3.8 is broad and does not apply to a particular "mode of computation" (e.g., deciding predicates [6, 7], computing functions [12, 22, 28], leader election [16, 31]). It does, however, assume a problem-specific notion of **valid** initial configurations.[9] We say a protocol is **additive** if the set of valid initial configurations is closed under addition. This rules out, for instance, protocols with an initial leader. Indeed, Corollary 3.9 is false if an initial leader is allowed, by applying Theorem 3.3 to let a leader-driven $O(1)$-message protocol simulate any $o(n)$-state open protocol that stably computes a non-semilinear predicate/function.[10]

A lower bound result in [21] shows that with an absolute space bound [11] of $o(\log n)$ states, their model is limited to only stably computing the semilinear predicates.[12] The core of their argument bounds the number of reachable memory states.

▶ **Theorem 3.7** ( [21]). *Let $s : \mathbb{N} \to \mathbb{N}$ and consider an additive, $s(n)$-state, open population protocol. Then either $s(n) = O(1)$ or $s(n) = \Omega(\log n)$.*

---

[9] For example, for leader election, all agents have the same initial state. For computation of predicates [6] or functions [12, 22], all agents represent "input" from a constant alphabet, with possibly an extra leader.

[10] For example, transitions $(i; \ell), (i; \ell) \to (i+1; \ell), (i+1; f)$ and $(j; \ell), (i; f) \to (j; \ell), (j; f)$, which starting from all agents in state $(1, \ell)$, give each agent the value $\lfloor \log n \rfloor$.

[11] The notion of a state bound has a few different meanings. By "absolute", we mean that $s(n)$ is the most number of states producible from any valid initial configuration of size $n$. Some uniform protocols (those without pre-programmed knowledge of $n$) have a space bound $s(n)$ that is only probabilistic, so memory usage can (with low probability) grow arbitrarily large in a fixed population; for example, see [18, 26, 32, 33] or Protocol 2.

[12] Theorem 14 of [21] states "$o(\log \log n)$" bits, which implies $o(\log n)$ states, though the converse does not hold. However, inspecting their proof reveals that the result holds up to $\log(n) - 1$ states.

As a corollary, if $s(n) = o(\log n)$, then $s(n)$ is in fact constant, reducing to the original $O(1)$-state model, which can only stably compute semilinear predicates [7]. We use a similar proof technique to show an exponentially stronger result in the model of $O(1)$ messages.

▶ **Theorem 3.8.** *Let $s : \mathbb{N} \to \mathbb{N}$ and consider an additive, $s(n)$-state, $O(1)$-message population protocol. Then either $s(n) = O(1)$ or $s(n) = \Omega(n)$.*

**Proof sketch.** A fixed population $\vec{i}_c$ suffices to produce any of the $O(1)$ messages. Consider a population $\vec{i}_n$ of size $n$. If $s(n) \neq O(1)$, then for some state $b$ not producible from $\vec{i}_n$, $b$ **is** producible by sending some message $m$ to a state $a$ producible from $\vec{i}_n$ (though $a$ and $m$ cannot appear *simultaneously* in a configuration reachable from $\vec{i}_n$). By combining $\vec{i}_n$ with $\vec{i}_c$, we have a population of size $n + O(1)$ that can produce $b$. Thus the number of producible states grows at least linearly with $n$. ◀

**Proof.** If $s(n) = O(1)$ we are done, so assume $s(n)$ grows without bound. Let $M(n)$ (respectively, $S(n)$) be the set of all messages (respectively, states) producible from a valid initial configuration of size $n$. Note $|M(n)| = O(1)$ and $|S(n)| = s(n)$. It suffices to show that for some constant $\epsilon > 0$ depending on the protocol, there are infinitely many $n$ such that $|S(n)| \geq \epsilon n$.

Let $c$ be the smallest population size $n$ such that $M(n)$ is the set of all messages $M$. We do not require all messages to be producible **simultaneously**, only that for each $m \in M(n)$, there is a valid initial configuration $\vec{i}_m$ such that $m$ can be produced from $\vec{i}_m$. Let $\epsilon = 1/c$. Inductively assume for some $n \in \mathbb{N}^+$ that $|S(n)| \geq \epsilon n$. Let $n' = n + c = n + 1/\epsilon$. It suffices to show that $|S(n')| \geq |S(n)| + 1 = \epsilon n + 1 = \epsilon n'$, i.e., a new state not in $S(n)$ is producible from some valid initial configuration of size $n'$.

There is some state $b \notin S(n)$ producible by an interaction of an agent in state $a \in S(n)$ with some message $m \in M$. Let $\vec{i}_n$ be a valid initial configuration of size $n$ from which $a$ is producible, and let $\vec{i}_c$ be a valid initial configuration of size $c$ from which $m$ is producible. Define $\vec{i}_{n'} = \vec{i}_n + \vec{i}_c$, which is valid because the protocol is additive. Since $a$ is producible from $\vec{i}_n$ and $m$ is producible from $\vec{i}_c$, $a$ and $m$ are simultaneously producible from $\vec{i}_{n'}$. By interacting the agent in state $a$ with the agent with message $m$, the state $b \notin S(n)$ is produced. Thus $|S(n')| \geq |S(n)| + 1$. ◀

Population protocols using $O(1)$ states compute only semilinear predicates [7], resulting in the following corollary. Since we require additivity of valid initial configurations, the corollary applies only to leaderless protocols.

▶ **Corollary 3.9.** *If a leaderless, $o(n)$-state, $O(1)$-message protocol stably computes a predicate $\phi$, then $\phi$ is semilinear.*

Corollary 3.9 is asymptotically tight by Observation 3.10.

▶ **Observation 3.10.** *For every $\epsilon > 0$, there is a leaderless, $(\epsilon n + O(1))$-state, $6$-message protocol stably computing a non-semilinear predicate.*

**Proof.** Let $c \in \mathbb{N}^+$ and consider the protocol where each agent's internal state is a natural number $k \in \mathbb{N}$, initially 1, representing a number of "balls." Each message $m \in \{0, 1, c\} \times \{L, F\}$ represents a number of balls to give away and a leader bit. Each state is a leader bit and a counter $k$. Agents conduct leader election by fratricide $(L, L \to L, F)$. The leaders will collect balls from only the followers, and only in units of $c$ balls. Thus all followers with counter $k \geq c$ display the message $m = (c, F)$, and only interact with a leader. In this interaction, the leader increments $k$ by $c$ and the follower decrements $k$ by $c$. This guarantees the leader's counter $k$ only actually uses values $\{1 + ic : i \in \mathbb{N}\}$. Finally, followers with counter $1 \leq k < c$ display the message $m = (1, F)$. If two agents with $m = (1, F)$ interact, the initiator gives one ball to the responder (i.e. one increments $k$, one decrements $k$).

It is straightforward to show that eventually this protocol will stabilize to a single leader with count "about $n$": $k = 1 + c \lfloor \frac{n-1}{c} \rfloor$. The sum of counts are clearly preserved. Followers with $k \geq c$ balls must eventually give all units of $c$ balls to the leader, who never decreases its count. While there are still $j \geq c$ total balls among the followers, eventually some follower will collect $c$ balls to give to the leader.

Notice that this protocol can only achieve counter values $k \in \{0, 1, \ldots, c, 1 + c, 1 + 2c, 1 + 3c, \ldots, 1 + c \lfloor \frac{n-1}{c} \rfloor\}$, thus this counter uses $\frac{n}{c} + O(1)$ states. Counting the 6 messages gives $\frac{n}{6c} + O(1)$ states.

Finally, the leader can compute some non-semilinear predicate of its count $k = 1 + c \lfloor \frac{n-1}{c} \rfloor$ (e.g., whether $\lfloor \frac{n-1}{c} \rfloor = 2^j$ is a power of two[13]), and use its message bit to tell the output to the rest of the population. (So a follower seeing message $m = (i, L)$ sets its output bit to $i$.) ◄

## 4  Timing Lemmas

The following technical lemmas are about the relationship between agents' local counts and the global number of interactions, and the time it takes to spread information by epidemic. They are used in the runtime analysis of our time efficient protocols.

### 4.1  Clock Drift Lemma

▶ **Lemma 4.1.** *Consider some interval of an execution of a population protocol with uniform random scheduling. Let $A_{is}$ be the indicator variable for the event that agent $i$ is one of the two agents that interact in step $s$ of this interval. Let $C_{it} = \sum_{s=1}^{t} A_{is}$ be the cumulative number of interactions involving agent $i$ during the first $t$ steps of the interval. Fix two agents $i$ and $j$, and let $\tau$ be the first time at which $C_{i\tau} + C_{j\tau} = m$. Then $\Pr[\max_{t \leq \tau}(C_{it} - C_{jt}) > b] \leq e^{-b^2/2m}$.*

**Proof.** Because only steps involving at least one of $i$ or $j$ change $C_{is}$ and $C_{js}$, we can restrict our attention to the sequence of steps $s_1, s_2, \ldots$ at which at least one of $i$ or $j$ interacts. Let $X_k = A_{it_k} - A_{jt_k}$; then $E[X_k \mid X_1, \ldots, X_{k-1}] = 0$ and the $X_k$ form a martingale difference sequence with $|X_k| \leq 1$. We also have that $C_{it_k} - C_{jt_k} = \sum_{\ell=1}^{k} X_\ell$, so $\max_{t \leq \tau}(C_{it} - C_{jt}) > b$ if and only if there is some $k$ with $t_k \leq \tau$ such that $\sum_{\ell=1}^{k} X_\ell > b$. Define the truncated martingale difference sequence $Y_k = X_k$ if $t_k \leq \tau$ and $\sum_{\ell=1}^{k-1} X_\ell \leq b$, and $Y_k = 0$ otherwise. Let $S_k = \sum_{\ell=1}^{k} Y_\ell$.

We have defined $S_k$ so that it tracks $C_{it_k} - C_{jt_k}$ until that quantity reaches $b + 1$ or $t_k$ reaches $\tau$, after which $S_k$ does not change. The condition $t_k = \tau$ occurs for some $k \leq m$, so if $C_{it_k} - C_{jt_k}$ reaches $b + 1$ before $t_k > \tau$, it must do so for some $k \leq m$, after which $S$ will not change, giving $S_m = S_k$. So $\Pr[\max_{t \leq \tau}(C_{it} - C_{jt}) > b] = \Pr[S_m > b] \leq e^{-b^2/2m}$, by the Azuma-Hoeffding inequality. ◄

---

[13] Our model does not directly count the memory required to compute the transition $\delta$. However, for this argument the Turing Machine would only need $O(1)$ bits of overhead, storing a counter $i$ to represent $1 + ic$, which is a power of two if and only if its binary expansion matches the regex $10^*$. Thus this asymptotic tightness on states holds even under a stricter state-complexity definition that counts space requirements of local computation.

▶ **Corollary 4.2.** *For any agents $i$ and $j$, and any $m$ and $b$,* $\Pr\left[\exists t : C_{it} = m \wedge C_{jt} > m + b\right] < e^{-b^2/(2m+b+1)}$.

**Proof.** If $C_{jt} - C_{it} > b$ when $C_{it} = m$, then there is some first time $s$ at which $C_{js} - C_{is} > b$. Because $s \leq t$, $C_{is} \leq C_{it} = m$, and because this is the first time at which $C_{js} - C_{is} > b$, we have $C_{js} = C_{is} + b + 1 \leq m + b + 1$. So $s$ is a time at which $C_{is} + C_{js} \leq 2m + b + 1$ with $C_{js} - C_{is} > b$. Now apply Lemma 4.1. ◂

## 4.2 Drift Fraction Lemma

Recall that $\mu$ units of time is defined as $\frac{n}{2} \cdot \mu$ interactions.

▶ **Lemma 4.3.** *Consider some set $S$ of agents and interval of length $T = \frac{n}{2} \cdot \mu$ interactions. Let $L \subseteq S$ be the subset of $S$ who have less than $\mu - l$ interactions during the interval. Then for $\epsilon_L = 2\sqrt{2\ln(n)/|S|} + \exp(-l^2/2\mu)$, $\Pr\left[|L| \leq \epsilon_L |S|\right] \geq 1 - 1/n^2$ and $\Pr\left[|L| = 0\right] \geq 1 - |S|\exp(-l^2/2\mu)$.*

*Likewise, let $H \subseteq S$ be the subset of $S$ who have more than $\mu + h$ interactions during the interval, where $h \leq \mu$. Then for $\epsilon_H = 2\sqrt{2\ln(n)/|S|} + \exp(-h^2/3\mu)$, $\Pr\left[|H| \leq \epsilon_H |S|\right] \geq 1 - 1/n^2$ and $\Pr\left[|H| = 0\right] \geq 1 - |S|\exp(-h^2/3\mu)$.*

**Proof.** For each agent $v$ and step $t$ of the interval, let $A_{v,t}$ be the indicator variable for the event that agent $v$ is one of the two agents that interact in step $t$ (with $\Pr\left[A_{v,t} = 1\right] = \frac{2}{n}$). For each agent $v$, let $L_v$, $H_v$ to be the indicator variables for the events that agent $v$ participates in fewer than $\mu - l$ interactions and more than $\mu + h$ interactions, respectively. Then $L = \{v \in S : L_v = 1\}$ and $H = \{v \in S : H_v = 1\}$, so $S' = L \cap H$.

Let $C_v = \sum_{t=1}^{T} A_{v,t}$ be the number of interactions in which agent $v$ participates, with $\mathrm{E}\left[C_v\right] = T \cdot \frac{2}{n} = \mu$. Since each step is independent, we can apply standard Chernoff bounds on the probability

$$\Pr\left[L_v = 1\right] = \Pr\left[C_v < \mu - l\right] = \Pr\left[C_v < \mu(1 - l/\mu)\right] \leq \exp\left(-(l/\mu)^2 \mu/2\right) = \exp\left(-\frac{l^2}{2\mu}\right).$$

Likewise, we can get an upper bound on the probability

$$\Pr\left[H_v = 1\right] = \Pr\left[C_v > \mu + h\right] = \Pr\left[C_v > \mu(1 + h/\mu)\right] \leq \exp\left(-(h/\mu)^2 \mu/3\right) = \exp\left(-\frac{h^2}{3\mu}\right).$$

Then $\Pr\left[L = 0\right] \geq 1 - |S|\exp\left(-\frac{l^2}{2\mu}\right)$ and $\Pr\left[H = 0\right] \geq 1 - |S|\exp\left(-\frac{h^2}{3\mu}\right)$ follow simply from the union bound over all $v \in S$. It is less straightforward to place high probability bounds on $|L|$ and $|H|$, because the indicator variables $L_v$ and $H_v$ are not independent. Intuitively, however, they have negative dependence, since if $L_v = 1$, that agent had a small number of interactions, making other agents more likely to have more. We formalize this intuition by showing that the joint distributions $\{L_v : v \in S\}$ and $\{H_v : v \in S\}$ are each **negatively associated**, which allows Chernoff bounds to be applied [30, Section 3.1].

First we show that for fixed $t$, the distribution $\{A_{v,t} : v \in [n]\}$ is negatively associated, since precisely two variables will have value 1 and the rest 0. Theorem 2.11 of [35] shows that all permutation distributions (random variables $X_1, \ldots, X_n$ whose values are a random permutation of $x_1, \ldots, x_n$) are negatively associated. The distribution $\{A_{v,t} : v \in [n]\}$ is a special case of a permutation distribution where the variables are a random permutation of $0, \ldots, 0, 1, 1$, and thus is negatively associated. Therefore the entire distribution $\{A_{v,t} : v \in [n], t \in [T]\}$ is negatively associated as the union of independent negatively associated variables (over the independent steps of the uniform scheduler).

Finally, the distribution $\{L_v \mid v \in S\}$ is a collection of monotone functions each defined on a disjoint subset of the distribution $\{A_{v,s} : v \in [n], s \in [T]\}$. This is precisely the property of disjoint monotone aggregation ( [30]), which shows that $\{L_v \mid v \in S\}$ is negatively associated. The same argument also holds for $\{H_v : v \in S\}$.

Now we will actually apply Chernoff bounds to the complements $|S \setminus L| = \sum_{v \in S}(1 - L_v)$ and $|S \setminus H| = \sum_{v \in S}(1 - H_v)$, which are justified by negative association. Then by linearity of expectation, $\sigma = \mathrm{E}\left[|S \setminus L|\right] \geq (1 - \exp(-l^2/2\mu))|S|$.

Now letting $\delta = 2\sqrt{2 \ln n / |S|}$ and $\epsilon_L = \delta + \exp(-l^2/2\mu)$, we have

$$\begin{aligned}
\Pr\left[|L| > \epsilon_L |S|\right] &= \Pr\left[|S \setminus L| < (1 - \epsilon_L)|S|)\right] \\
&\leq \Pr\left[|S \setminus L| < (1 - \delta)(1 - \exp(-l^2/2\mu))|S|\right] \\
&\leq \Pr\left[|S \setminus L| < (1 - \delta)\sigma\right],
\end{aligned}$$

and we can apply the Chernoff bound to get

$$\Pr\left[|S \setminus L| < (1 - \delta)\sigma)\right] \leq \exp\left(-\delta^2 \sigma / 2\right) \leq \exp\left(-\delta^2 |S|/4\right) \leq 1/n^2,$$

where we also assumed $\sigma \geq |S|/2$ since $\Pr\left[C_v \geq \mu - l\right] > 1/2$. Thus $\Pr\left[|L| \leq \epsilon_L |S|\right] \geq 1 - 1/n^2$.

The same argument, for $\epsilon_H = \delta + \exp(-h^2/3\mu)$ also shows that $\Pr\left[|H| \leq \epsilon_H |S|\right] \geq 1 - 1/n^2$. ◀

## 4.3 Epidemics

Recall an **epidemic** process in a population protocol starts with a single agent **infected** ($i$) and all others **susceptible** ($s$), and every encounter between an infected and uninfected agent causes the latter to become infected (i.e., the transition $i, s \to i, i$). Let $H_k = \sum_{i=1}^{k} \frac{1}{i} \sim \ln n$ be the $k$'th harmonic number. The following lemma, due to [40], was proved in its current form in [20].

▶ **Lemma 4.4** ( [40]). *Starting from a population of size $n$ with a single infected agent, let $T_n$ be the number of interactions until all agents are infected. Then $\mathrm{E}\left[T_n\right] = (n-1)H_{n-1} \sim n \ln n$, and for $n \geq 8$ and $\delta \geq 0$,*

$$\Pr\left[T_n > (1 + \delta) \mathrm{E}\left[T_n\right]\right] \leq 2.5 \ln(n) \cdot n^{-2\delta}.$$

The following lemma generalizes to a susceptible subset of the population, and considers a symmetric middle interval of an epidemic process, to be used in the analysis of Protocol 2.

▶ **Lemma 4.5.** *Consider the two way epidemic process starting from a configuration of $n$ agents with $a = \alpha n$ infected agents and $b = \gamma n \geq a$ susceptible agents (and $n - a - b$ agents not participating). Let $T$ be the number of interactions to reach $b + 1$ infected agents (with $a - 1$ susceptible agents left). Then $T \leq \frac{5}{\gamma} n \ln\left(\frac{\gamma}{\alpha}\right)$ with probability $1 - (\gamma n)^{-2}$.*

**Proof.** The number of infected agents $i$ will increase monotonically from $a$ to $b + 1$. At each stage, when there are $i$ infected agents and $b + a - i$ susceptible agents, there are $i(b + a - i)$ pairs of agents whose interaction increases the number of infected agents, so the next interaction is one of these with probability $p_i = \frac{i(b+a-i)}{\binom{n}{2}}$. Thus $T = \sum_{i=a}^{b} G_i$ where $G_i$ is a geometric random variable with parameter $p_i$. Then

$$\mu = E(T) = \sum_{i=a}^{b} \frac{1}{p_i} = \binom{n}{2} \sum_{i=a}^{b} \frac{1}{i(b+a-i)} = \frac{n(n-1)}{2(b+a)} \sum_{i=a}^{b} \frac{1}{i} + \frac{1}{b+a-i} = \frac{n(n-1)}{(b+a)}(H_b - H_{a-1})$$

where $H_n = \sum_{i=1}^{n} \frac{1}{i}$ is the $n$th Harmonic number. Then

$$\mu \approx \frac{n}{\alpha + \gamma} \ln\left(\frac{b}{a}\right) = \frac{n}{\alpha + \gamma} \ln\left(\frac{\gamma}{\alpha}\right)$$

and because $\gamma \geq \alpha > 0$ we can bound

$$\frac{n}{2\gamma} \ln\left(\frac{\gamma}{\alpha}\right) \leq \mu \leq \frac{n}{\gamma} \ln\left(\frac{\gamma}{\alpha}\right)$$

Then by Theorem 2.1 of [34], we have the large deviation bound

$$\Pr[T \geq \lambda \mu] \leq \exp\left(-p_* \mu (\lambda - 1 - \ln \lambda)\right)$$

where $p_* = \min_i p_i = \frac{ab}{\binom{n}{2}} \approx 2\alpha\gamma$.

Letting $\lambda = 5$, $\lambda - 1 - \ln \lambda > 2$, this gives

$$\Pr\left[T \geq \frac{5}{\gamma} n \ln\left(\frac{\gamma}{\alpha}\right)\right] \leq \Pr[T \geq 5\mu] \leq \exp\left(-2\alpha n \ln\left(\frac{\gamma}{\alpha}\right)\right).$$

Now observe that $\alpha \ln(1/\alpha)$ is increasing for all $\alpha \in (0, 1)$. Since $\alpha \geq 1/n$, we can take this minimum value to get

$$\Pr\left[T \geq \frac{5}{\gamma} n \ln\left(\frac{\gamma}{\alpha}\right)\right] \leq \exp\left(-2 \ln(\gamma n)\right) = (\gamma n)^{-2}.$$

◀

## 5 Computability with polylogarithmic time complexity

In this section we study $O(1)$-message population protocols when the goal is "fast" computation (polylog($n$) time).

### 5.1 High-probability junta election using 1-bit messages

In this section, we describe a uniform protocol using 1-bit messages that, with high probability, elects a "junta" of $O(\sqrt{n})$ agents in polylogarithmic time. The protocol also lets each agent compute an integer $k \in \mathbb{N}^+$ that is the same for all agents and is one of $\lfloor \log \log n \rfloor$, $\lceil \log \log n \rceil$, or $\lceil \log \log n \rceil + 1$. Thus $2^k$ is an estimate of $\log n$ within a multiplicative factor 2.

Furthermore, *JuntaElection* is composable, in that we can use the protocol as a black box to initialize other protocols that require either a junta for a phase clock, or an approximation of $\log n$ (e.g. for a leaderless phase clock). Thus, for any nonuniform protocol that requires $k$-bit messages, we can compose it with our *JuntaElection* protocol and achieve a uniform protocol that uses $(k + 1)$-bit messages with an additive time overhead of $O(\log^2 n)$. For example, we can compose the *JuntaElection* protocol with the the leader election protocol of [32] using $\frac{1}{2}$-coin flips to convert the $O(\sqrt{n})$-size junta to size 1, i.e., elect a unique leader, in expected $O(\log^2 n)$ time and $O(1)$ messages, or with majority protocols that use $O(1)$ messages for doubling/cancelling phases, synchronized by the junta-driven phase clock [14].

Our protocol has a positive probability of failure. It is an open question if there exists an $O(1)$-message protocol that can stably (i.e., with probability 1) approximate $\log n$ or elect a junta of size $n^\epsilon$ for some $0 < \epsilon < 1$ in sublinear stabilization time.

**Protocol 2** *JuntaElection*(Agent $v$ seeing message $m$)

**1 initial state of agent:**
**2** $GeometricRV \leftarrow \frac{1}{2}$-geometric random variable (used for estimating level)
**3** $v.\text{level} \leftarrow \lceil \log(GeometricRV) \rceil$ ;                                   // $v.\text{level} \in \mathbb{N}^+$
**4** $v.\text{count} \leftarrow 0$
**5** $v.\text{inJunta} \leftarrow True$

**6 if** $v.\text{count} = d_i$ **and** $v.\text{level} \leq i$ **then**
**7**  | **if** $m = \text{Go}$ **then**
**8**  |  | $v.\text{count} \leftarrow v.\text{count} + 1$
**9 else**
**10**  | $v.\text{count} \leftarrow v.\text{count} + 1$;
**11 if** $v.\text{count} \in G_i$ **and** $v.\text{level} \leq i$ **then**
**12**  | $v.\text{message} \leftarrow \text{Go}$
**13 else if** $v.\text{count} \in R_i$ **and** $v.\text{level} \leq i$ **then**
**14**  | $v.\text{message} \leftarrow \text{Stop}$
**15 if** $v.\text{count} \in [G_i \cup R_i]$ **and** $v.\text{level} > i$ **then**
**16**  | $v.\text{message} \leftarrow \text{Go}$
**17 if** $v.\text{count} = d_i$ **then**
**18**  | $lognEstimation \leftarrow 2^i$
**19**  | $v.\text{inJunta} \leftarrow v.\text{level} \geq i$

### 5.1.1 High-level description of protocol

The protocol is described formally in Protocol 2. Intuitively, it works as follows. Most other leader/junta election protocols generate an **id**, where the agents generating the maximum id are the junta. In our protocol, we also generate an id (called **level**), but $O(1)$ messages prevent direct communication of levels, so we employ a timing-based strategy for agents to learn the maximum level. The message consists of a single bit, taking values Go and Stop.

Each agent initially generates a local geometric random variable $G$ (number of fair coin flips until the first heads, i.e., an immediate heads results in $G = 1$) and computes its **level** as $\lceil \log G \rceil$. (We can also use synthetic coin techniques [1] to simulate fair coin flips and increment their level from $i$ to $i + 1$ as they flip $2^i$ consecutive tails.)

We define consecutive disjoint intervals $G_0, R_0, G_1, R_1, \ldots \subset \mathbb{N}$ (**green** and **red**) partitioning the natural number line. We call $R_i$'s last element $d_i = \max R_i$ a **door**. (See Figure 1, formal definition below.) Each agent keeps a local counter, initially 0, that is incremented on some interactions. An agent is **in round** $i$ if its counter is in $G_i \cup R_i$. The goal is to get every agent to count up until the round equal to the maximum level $k$ generated by any agent and stop its counter at $d_k$. An agent with level $l$ in round $i$ is **eager** if $i < l$ and **cautious** otherwise. Intuitively, eager agents race through doors until their own level, telling all other agents to keep going, but become cautious at and beyond their own level, advancing past a door into the next round only if another agent tells them to do so (via a message $m = \text{Go}$). More formally, an eager agent always sends a message of Go and increments its counter on every interaction. A cautious agent sends message Go if and only if its counter is in $G_i$ for some $i$, increments its counter on every interaction in $G_i \cup R_i \setminus \{d_i\}$ unconditionally, and increments its counter beyond $d_i$ if and only if the other agent's message is Go. Agents drop out of the junta when they leave their own level, so (assuming no agent leaves the maximum level) those who generated the maximum level are the eventual junta.

To formally defined the intervals, let $c \in \mathbb{N}^+$. Each $G_i$, with $|G_i| = c4^i$, is called a **green** interval, $R_i$, with $|R_i| = \frac{3c}{2}4^i$, a **red** interval. Note that $d_i = \sum_{j=0}^{i-1}(|G_j| + |R_j|) = c\left(1 + \frac{3}{2}\right)\frac{4^i - 1}{4 - 1} < \frac{5c}{6}4^i = \frac{5}{6}|G_i|$, so $|G_i|$ is larger by a constant multiplicative factor than the union of all the previous intervals. The max level $k$ is $\Theta(\log \log n)$ with high probability, so its corresponding interval $|G_k| = \Theta(4^{\log \log n}) = \Theta(\log^2 n)$. Thus, with high probability, the agents will use $O(\log^2 n)$ states for their counters and stop at the door $d_k$ after $O(\log^2 n)$ time.

To compose *JuntaElection* with a downstream protocol $P$, agents can simply restart $P$ whenever they move beyond a $d_i$, and then wait to start simulating $P$ until they reach the next $d_{i+1}$ (Restarting is a common technique in distributed computing for composition and is not original to this paper, e.g., [32].) In the early stages of *JuntaElection*, the downstream protocol gets restarted many times, but eventually, all agents will move past $d_{k-1}$, after which they will restart the downstream protocol for the last time. The agents will all simultaneously be in the last interval $G_k \cup R_k$ before stopping at $d_k$. Thus all simulated downstream interactions of $P$ will be between agents that agree on $k$.

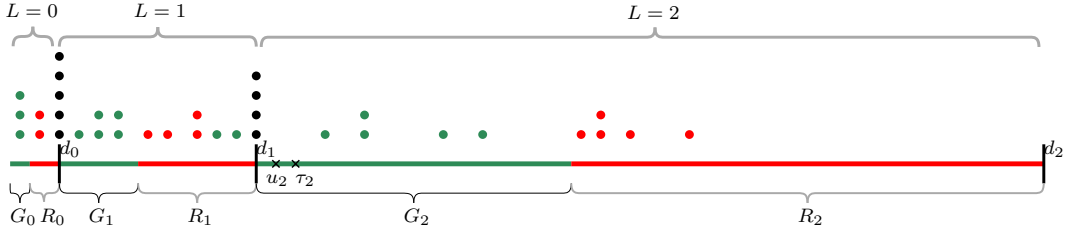

**Figure 1** Agents, represented as dots, increment their counters through the $G_0, R_0, G_1, R_1, G_2, R_2$ intervals. Agents in green intervals or any interval before their own level have message `Go`. Agents in red intervals at their own level or later have message `Stop`. At the end of a red interval (the door $d_i$, shown with black horizontal line) at their own level or later, the agents (black dots) wait to increment their counter until they see a message `Go`. The special times marked $u_i, \tau_i$ are used in proving Lemma 5.4.

▶ **Theorem 5.1.** *With probability $1 - O(1/n)$, Protocol 2 uses $O(\log^2 n)$ states and elects a junta of size $O(\sqrt{n})$ in $O(\log^2 n)$ time, after which $v$.count $= d_k$ for all agents $v$, where $k \in \{\lfloor \log \log n \rfloor, \lceil \log \log n \rceil, \lceil \log \log n \rceil + 1\}$.*

Theorem 5.1 is proven formally via Lemmas 5.2, 5.3, 5.4, 5.5, 5.6 in Subsection 5.2.

**Proof sketch.** We must show the agents remain synchronized. By the time the interval lengths are $\Omega(\log n)$, we could argue that the number of interactions of each agent are tightly concentrated enough for agents to by synchronized in the same interval. However, the main challenge is how to reason about agents that might be stuck behind at a door.

Our argument shows that a constant fraction $n/4$ of agents stay synchronized in each green interval, up until near the max level (Lemma 5.4). Then, we argue that during the later green intervals, straggler agents are able to catch up, because they have a constant probability of passing through each door and the length of the green interval is more than the sum of all previous intervals. We then show the entire population in synchronized within the last few intervals (Lemma 5.5). Thus all agents will have a `Stop` message when the population reaches the final door $d_k$, and the agents will stop their counters at $d_k$. ◀

Our proof techniques require setting $|G_i| = 700 \cdot 4^i$. However, simulation results (Figure 2) show successful convergence when $|G_i| = 16 \cdot 2^i$. Scaling the intervals this way would let $|G_M| = \Theta(\log n)$, so the protocol would take $O(\log n)$ time and $O(\log n)$ internal states.

## 5.2 Proof of Theorem 5.1

Theorem 5.1 follows from Lemmas 5.2, 5.3, 5.4, 5.5, 5.6, proven in this subsection.

### 5.2.1 Distribution of levels

▶ **Lemma 5.2.** *Let $n \in \mathbb{N}^+$ and consider $n$ i.i.d. geometric random variables $V_1, \ldots, V_n$. Let $i \in \mathbb{N}^+$, $E_i = |\{j \mid \lceil \log V_j \rceil = i\}|$. Let $0 < \delta < 1$. Let $\mu = n\left(2^{-2^{i-1}} - 2^{-2^i}\right)$. Then $\Pr\left[(1-\delta)\mu < E_i < (1+\delta)\mu\right] > 1 - 2 \cdot \exp\left(-n\delta^2 2^{-1-2^{i-1}}/3\right)$.*

**Proof.** Let $G$ be a geometric random variable, so for each $a \in \mathbb{N}$, $\Pr[G > a] = 2^{-a}$. Then for each $i \in \mathbb{N}$, $\Pr[\lceil \log G \rceil > i] = \Pr\left[G > 2^i\right] = 2^{-2^i}$. Then $\Pr[\lceil \log G \rceil = 0] = 1/2$ and for $i > 0$, $\Pr[\lceil \log G \rceil = i] = \Pr[\lceil \log G \rceil > i-1] - \Pr[\lceil \log G \rceil > i] = 2^{-2^{i-1}} - 2^{-2^i}$. Note that for all $a \in \mathbb{N}^+$, $2^{-a} - 2^{-2a} \geq 2^{-1-a}$, so $2^{-2^{i-1}} - 2^{-2^i} \geq 2^{-1-2^{i-1}}$.

Now consider $n$ i.i.d. variables $V_1, \ldots, V_n$. For each $j \in \{1, \ldots, n\}$, let $I_j$ be the indicator for the event $V_j \in L_i$. Let $p = 2^{-2^{i-1}} - 2^{-2^i} \geq 2^{-1-2^{i-1}}$, noting $\Pr[I_j = 1] = p$ and $\sum_{j=1}^n I_j = E_i$, with $\mu = \mathrm{E}[E_i] = np$. Since $p > 2^{-1-2^{i-1}}$, it follows that $\mu \geq n \cdot 2^{-1-2^{i-1}}$. By independence of the $V_j$'s and the Chernoff bound,

$$\Pr\left[E_i < (1-\delta)\mu \text{ or } E_i > (1+\delta)\mu\right] < 2 \cdot \exp(-\delta^2 \mu/3) < 2 \cdot \exp(-\delta^2 n 2^{-1-2^{i-1}}/3). \quad \blacktriangleleft$$

Letting $\delta = 1/2$, and noting $n/2^{2^{i-1}+1} < \mu < n/2^{2^{i-1}-1}$, gives the following corollary.

▶ **Corollary 5.3.** *Let $n \in \mathbb{N}^+$ and consider $n$ i.i.d. geometric random variables $V_1, \ldots, V_n$. For each $i \in \mathbb{N}^+$, let $E_i = |\{j \mid \lceil \log V_j \rceil = i\}|$. Then $\Pr\left[n/2^{2^{i-1}+2} < E_i < n/2^{2^{i-1}}\right] > 1 - 2 \cdot \exp\left(-n 2^{-5-2^{i-1}}\right)$.*

Note that Corollary 5.3 is useful as long as $i \leq \log \log n$. There is a $\Theta(1)$ failure probability when $i = 1 + \log \log n$, and a very large failure probability when $i \geq 2 + \log \log n$. But for $i = \log \log n$ (and smaller), the failure probability is at most $2 \exp\left(-n^{1/2}/32\right)$. Of course, $i$ is an integer and $\log \log n$ in general is not; nevertheless, with appropriate rounding we conclude that $k = \max_{j \in \{1, \ldots, n\}} \lceil \log V_j \rceil$ is very likely to be $\lfloor \log \log n \rfloor$, $\lceil \log \log n \rceil$, or $\lceil \log \log n \rceil + 1$.

In the following we use the fact that $E_i$ is the number of agents choosing exactly level $i$.

For any field field of an agent $v$ and any $t \in \mathbb{N}$, let $v.\text{field}(t)$ denote the value of field in agent $v$ at time $t$ ($\frac{n}{2}t$ interactions). Write $v.\text{field}$ when the time is clear from context (or $v.\text{field}$ is constant over time, e.g. $v.\text{level}$).

Define $u_i, \tau_i$ to be the points $\frac{1}{16}$ and $\frac{1}{8}$ of the way through interval $G_i$ (see Figure 1). Thus $u_i = d_{i-1} + \frac{c}{16}4^i$ and $\tau_i = d_{i-1} + \frac{c}{8}4^i$. (Recall we have the bound $d_{i-1} < \frac{5}{6}4^i$.) At time $\tau_i$, the average number of interactions is $\tau_i$, and we hope for most agents' counters to also be near $\tau_i$. $S_i$ denotes the cautious agents that, at time $\tau_i$, are synchronized with counters in the interval $G_i$. The following lemma shows that a constant fraction of the population are in $S_i$:

▶ **Lemma 5.4.** *Let $i \in \{0, 1, \ldots, \lfloor \log \log n \rfloor - 1\}$. Let $S_i$ be the set of agents $v$ such that $v.\text{level} \leq i$ (cautious by round $i$) and $v.\text{count}(\tau_i) \in G_i$. Then $\Pr\left[|S_i| \geq n/4\right] > 1 - O\left(\frac{\log \log n}{n^2}\right)$.*

**Proof.** We prove this by induction on $i$.
<u>Base case.</u> $S_0$ is the set of agents $v$ with $v.\text{level} = 0$ and $v.\text{count}(\tau_0) \in G_0$. Then $\mathrm{E}[E_0] = n/2$, and taking $\delta = 1/4$ in Lemma 5.2, we get

$$\Pr\left[E_0 < 3n/8\right] = \Pr\left[E_0 > (1-\delta)\mu\right] \leq 2 \cdot \exp(-n 2^{-5.5}/3).$$

We now show that of these level 0 agents, we only lose an additional fraction of 1/3 due to drift in the first $\tau \cdot n/2$ interactions ($\tau$ units of time). We apply Lemma 4.3, with $S$ as the agents at level 0, $\mu = \tau_0 = c/8$, and $\mu + h = \max G_i = c$. Then $S' = S_0$, and the error fraction

$$\epsilon_H = 2\sqrt{2\ln n/|S|} + \exp(-h^2/3\mu) \leq 2\sqrt{2\ln n/(3n/8)} + \exp(-49c/24) \leq 1/3,$$

for sufficiently large values of $n$. Then by Lemma 4.3, $|S_0| \geq (1 - 1/3)(3n/8) = n/4$ with probability $1 - 1/n^2$.

<u>Inductive case.</u> Assume $|S_i| \geq n/4$, recalling that for all $v \in S_i$, $v.\text{level} \leq i$ and $v.\text{count}(\tau_i) \in G_i$. We will first wait until time $u_{i+1}$, and consider the agents from $S_i$ and also those at level $i+1$ that have at least made it to the door $d_i < u_{i+1}$.

Let $A_{i+1}$ be the set of agents $v$ with $v.\text{level} = i+1$ and $v.\text{count}(u_{i+1}) \geq d_i$. Let $B_{i+1}$ be the set of agents $v \in S_i$ with $v.\text{count}(u_{i+1}) \geq d_i$.

Intuitively, the agents in $B_{i+1}$ have had enough interactions to at least be at the door $d_i$, but could be stuck waiting to see the signal. The agents in $A_{i+1}$ are broadcasting a signal and moving the agents in $B_{i+1}$ through the door. This process will be stochastically dominated by a section of an epidemic process. We must ensure $|A_{i+1}| + |B_{i+1}| > n/4$ so that we can wait for $|S_{i+1}| \geq n/4$ agents to finish this epidemic.

First, we must bound the size $|A_{i+1}|$. Let $E_{i+1}$ be the set of agents $v$ with $v.\text{level} = i+1$. Then by Corollary 5.3, $\Pr\left[|E_{i+1}| > n/2^{2+2^i}\right] > 1 - \exp(-n2^{-5-2^i})$. Since $i+1 \leq \lfloor \log \log n \rfloor - 1$, we have

$$|E_{i+1}| > n/2^{2+2^{\log\log n - 2}} = n/(4 \cdot 2^{\log n/4}) = \frac{1}{4}n^{3/4}$$

Now $A_{i+1}$ is the subset of $E_{i+1}$ that have at least $d_i$ interactions between time 0 and time $u_{i+1}$. We apply Lemma 4.3 with $S = E_{i+1}$, $l = u_{i+1} - d_i = \frac{c}{16}4^{i+1}$, and $\mu = u_{i+1} = d_i + \frac{c}{4}4^{i+1} < \left(\frac{5}{6} + \frac{1}{4}\right)c4^{i+1} < 18l$. Then we can use a simple upper bound for the fraction

$$\epsilon_L = 2\sqrt{2\ln n/|S|} + \exp(-l^2/2\mu) \leq 4\sqrt{8\ln n/n^{3/4}} + \exp\left(-\frac{c}{16 \cdot 18}4^{i+1}\right) \leq 1/2$$

as long as $c \geq 72$, for sufficiently large values of $n$. Then Lemma 4.3 will give that $|A_{i+1}| > (1 - \frac{1}{2})|E_{i+1}| > n/2^{3+2^i}$ with probability $1 - 1/n^2$.

Next we must bound the size $|B_{i+1}|$ again using Lemma 4.3. $S = S_i$, since we are starting from the agents in $S_i$ at time $\tau_i$ (who have $v.\text{count}(\tau_i) < d_{i-1}$). We will consider the drift during the interval between time $\tau_i$ and $u_{i+1}$, so

$$\mu = u_{i+1} - \tau_i = c\left(\frac{7}{8}4^i + \frac{3}{2}4^i + \frac{1}{16}4^{i+1}\right) = \frac{21c}{8} \cdot 4^i.$$

The agents in $B_{i+1}$ must have $v.\text{count}(u_{i+1}) \geq d_i$, meaning they have at least $\mu - l = d_i - d_{i-1} = c(1 + \frac{3}{2})4^i$ interactions.

Then $l = \frac{c}{8}4^i$, and Lemma 4.3 gives that $|B_{i+1}| \geq (1-\epsilon_L)|S_i| \geq \frac{n}{4}(1-\epsilon_B)$ with probability $1 - 1/n^2$, where

$$\epsilon_B = \epsilon_L = 2\sqrt{2\ln n/|S|} + \exp(-l^2/2\mu) \leq 4\sqrt{2\ln n/n} + \exp\left(-\frac{c}{336}4^i\right).$$

Now at time $u_{i+1}$ we have $|A_{i+1}| + |B_{i+1}|$ agents that are at least at door $d_i$. Agents from $B_{i+1}$ might be stuck at the door, but they will advance past as soon as they encounter

an agent from $A_{i+1}$ or another agent from $B_{i+1}$ that has already past the door. Thus this looks like an epidemic process where $|A_{i+1}| + |B_{i+1}|$ agents are participating, and we start with at least $|A_{i+1}|$ infected agents. We will wait $\tau_{i+1} - u_{i+1}$ time and hope to reach at least $n/4$ infected agents.

However, there is the added complication that agents might drift past $G_{i+1}$, so they can't get counted in $S_{i+1}$ and will no longer be acting as an infected agent in the epidemic. We will again use Lemma 4.3 to bound the count $|D|$ of any agents that have more than $\max G_{i+1}$ interactions by time $\tau_{i+1}$. We use $\mu = \tau_{i+1} = d_i + \frac{c}{8}4^{i+1} < c(\frac{5}{6} + \frac{1}{8})4^{i+1}$, $h = \max G_{i+1} - \tau_{i+1} = \frac{7c}{8}4^{i+1}$, so $h^2/3\mu > \frac{49}{184}c4^{i+1} > c4^i$. We will consider the worst case for the size of $|D|$, with $|S| = n$ agents possible to drift. Lemma 4.3 gives that $|D| \leq \epsilon_D n$, where

$$\epsilon_D = 2\sqrt{2\ln n/|S|} + \exp(-h^2/3\mu) \leq 2\sqrt{2\ln n/n} + \exp\left(-c4^i\right)$$

Now we will make a worst case assumption that all drifted agents come from the initially infected agents $A_{i+1}$, and argue about an epidemic starting from $|A_{i+1}| - |D|$ infected agents with $|A_{i+1}| + |B_{i+1}| - |D|$ agents participating.

We will use the bound

$$|A_{i+1}| \geq \frac{1}{2}|E_{i+1}| = \frac{1}{4}|E_{i+1}| + \frac{1}{8}|E_{i+1}| + \frac{1}{8}|E_{i+1}| \geq \frac{1}{16}n^{3/4} + n/2^{5+2^i} + n/2^{5+2^i},$$

broken up with two terms that will dominate each of the terms in $\epsilon_B$ and $\epsilon_D$.

Then we can bound

$$|A_{i+1}| + |B_{i+1}| - |D| \geq \frac{1}{16}n^{3/4} + n/2^{5+2^i} + n/2^{5+2^i} + \frac{n}{4}\left(1 - 4\sqrt{2\ln n/n} - \exp\left(-\frac{c}{336}4^i\right)\right)$$
$$- n\left(2\sqrt{2\ln n/n} + \exp\left(-c4^i\right)\right)$$
$$\geq \frac{n}{4} + n/2^{5+2^i} + \left(\frac{1}{16}n^{3/4} - 16\sqrt{2n\ln n} - 2\sqrt{2n\ln n}\right)$$
$$+ n\left(2^{-5-2^i} - \frac{1}{4}\exp\left(-\frac{c}{336}4^i\right) - \exp(-c4^i)\right)$$
$$\geq \frac{n}{4} + n/2^{5+2^i} + 0 + 0$$

for sufficiently large values of $n$ (since $\sqrt{n\ln n} = o(n^{3/4})$), and for $c \geq 700$ (since the rightmost difference is minimized at $i = 0$ at positive for $c \geq 700$).

These calculations also show that $|A_{i+1}| - |D| \geq n/2^{5+2^i}$. Thus, the true process will be stochastically dominated by a two-way epidemic, starting from $a = n/2^{5+2^i}$ infected agents and $n/4$ susceptible agents. Recall we are waiting $\tau_{i+1} - u_{i+1}$ time, which is $\frac{n}{2} \cdot \frac{c}{16}4^{i+1} = n \cdot \frac{c}{8}4^i$ interactions.

Now we can apply Lemma 4.5 (with $\alpha = 2^{-5-2^i}$ and $\gamma = 1/4$) to conclude this process will reach $s + 1 > n/4$ infected agents after $T$ interactions, where

$$T \leq \frac{5}{\gamma}n\ln\left(\frac{\gamma}{\alpha}\right) = 20n\ln\left(2^{3+2^i}\right) = n \cdot 20\ln 2 \cdot (3 + 2^i) \leq n \cdot \frac{c}{8}4^i$$

with probability $1 - (\gamma n)^{-2} = 1 - 16/n^2$ (as long as $c \geq 333$).

Thus by time $\tau_{i+1}$, at least $n/4$ agents have passed the door $d_i$ without leave $G_{i+1}$. Therefore we have showed $|S_{i+1}| \geq n/4$ with probability $1 - O(1/n^2)$, completing the inductive case.

Note that we considered $i < \log\log n$ levels, where each of these inductive steps added an error probability $O(1/n^2)$. Therefore we have $\Pr[|S_i| \geq n/4] > 1 - O\left(\frac{\log\log n}{n^2}\right)$. ◂

▶ **Lemma 5.5.** *Let $i \in \{\lfloor \log \log n \rfloor - 1, \ldots, k\}$, where $k = \max_v v.\mathsf{level}$. Then with probability $1 - O(1/n)$, there is some time $t$ such that $v.\mathsf{count}(t) \in G_i$ for all agents $v$, and also some time $t$ such that $v.\mathsf{count}(t) \in R_i$ for all agents $v$.*

**Proof.** Notice that $i \geq (\log \log n) - 2$, so $4^i > \frac{1}{16} \log^2 n$, and for any constant $a > 0$, we have $\exp(-a 4^i) < \exp(-\frac{a}{16} \log^2 n) < 1/n^2$ for sufficiently large $n$. Thus we will find that all error terms of the form $\exp(-l^2/3\mu)$ and $\exp(-h^2/3\mu)$ from Lemma 4.3 will now be small enough that we can use the union bound result and conclude $|L| = 0$ and $|H| = 0$.

First we consider $i = \lfloor \log \log n \rfloor - 1$. By Lemma 5.4, $|S_i| \geq n/4$, meaning at time $\tau_i$, there are at least $n/4$ agents at level $\leq i$ in $G_i$.

Now let $\Omega = d_{i-1} + \frac{47c}{48} 4^i$ be the point $\frac{47}{48}$ of the way through the interval $G_i$. Let $D$ be the set of agents that have more than $\max G_i$ interactions by time $\Omega$. We apply Lemma 4.3 with $|S| = n$, $\mu = \Omega < (1 + \frac{5}{6})c 4^i < 2c 4^i$ and $\mu + h = \max G_i$, so $h = \frac{c}{48} 4^i$. Now we have $h^2/3\mu = \Theta(4^i)$, so as observed above $\exp(-h^2/3\mu) < 1/n^2$ and we can conclude that $|D| = 0$ with probability $1 - O(1/n)$.

Now we consider the time between $\tau_i$ and $\Omega$. Since no agent has had more than $\max G_i$ interactions, there are at least $n/4$ agents in $G_i$ during this entire interval. We can now show that every agent will enter $G_i$ by the time $\Omega$. In the worst case, an agent $v$ could still have $v.\mathsf{count}(\tau_i) = 0$. If $v$ has an interaction while at a door during this interval, the chance of increasing its counter is at least $\frac{n}{4}/n = 1/4$. The probability of taking more than $8 \ln n$ interactions to pass a door is at most $(1 - 1/4)^{8 \ln n} < \exp -\frac{8}{4} \ln n = 1/n^2$. The probability of taking more than $8 \ln n \log \log n$ interactions to pass all doors is then at most $\log \log n/n^2$. This number of interactions is negligible compared to $d_i = \Theta(4^i) = \Theta(\log^2 n)$.

Now we are waiting
$$\Omega - \tau_i = \left(\frac{7}{8} - \frac{1}{48}\right)|G_i| > \frac{5}{6}|G_i| > d_i$$
time and need to have $d_i + o(\log^2 n)$ interactions. Thus we can again apply Lemma 4.3, where again $\mu = \Theta(4^i)$ and $l = \Theta(4^i)$. This will show that with probability $1 - O(1/n^2)$, every agent has enough interactions between $\tau_i$ and $\Omega$ to reach the interval $G_i$.

We have now shown that $v.\mathsf{count}(\Omega) \in G_i$ for all $v$. Then the number of interactions for an agent to enter $R_i$ is at most $|G_i| = c 4^i$ and the number of interactions to reach $d_i$ is at least $|R_i| = \frac{3c}{2} 4^i$, so we can find some time $t \in R_i$ when Lemma 4.3 will give that with probability $1 - O(1/n^2)$, every agent will have at least enough interactions to enter $R_i$ but not enough interactions to reach $d_i$.

While $i < k$, there will be at least one agent at level $i + 1$. We can then make the same epidemic argument as in the proof of Lemma 5.4. Now, we can assume in the worst case we have an epidemic that starts with one agent that must reach the whole population. By either Lemma 4.4 or Lemma 4.5, this takes $O(\log n)$ time ($O(n \log n)$ interactions), with probability $1 - 1/n^2$. This is now negligible compared to the $\Theta(\log^2 n)$ time during each green interval.

Thus we can inductively claim that for all intervals $G_{i+1}, R_{i+1}, \ldots, G_M, R_M$ there is some time $t$ when all agents are synchronized with counts in that interval. ◂

Finally, we establish the state complexity bound for Protocol 2 stated in Theorem 5.1.

▶ **Lemma 5.6.** *With probability $1 - O(1/n)$, Protocol 2 uses $O(\log^2 n)$ internal states.*

**Proof.** Note the space is dominated by the field $\mathsf{count} = O(\log^2 n)$ with probability $1 - O(1/n)$. All the intervals are properties of the transition function $\delta$ not the state. Finally, as written holding the level would take $O(\log \log n)$ state overhead, but we could actually compute the level on the fly and only need to track if we are still eager or cautious in each interval. ◂

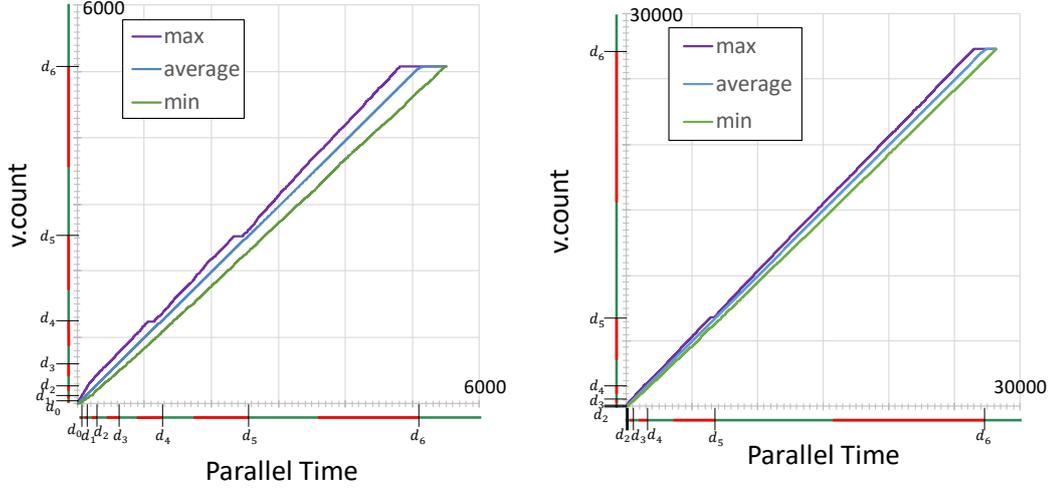

**(a)** $|G_i| = 16 \cdot 2^i$, $|R_i| = 24 \cdot 2^i$

**(b)** $|G_i| = 2 \cdot 4^i$, $|R_i| = 3 \cdot 4^i$

**Figure 2** Data from two simulations on $n = 2 \cdot 10^7$ agents, with different values of $|G_i|$ and $|R_i|$. Both yielded maximum level $k = 6$. The horizontal axis shows parallel time ($\frac{n}{2}$-interactions). The vertical axis represents the value $v$.count for agents (summarized for the whole population by min, max, and average). The purple line, the orange line, and the green line are respectively showing the maximum, average, and minimum $v$.count for agents. As shown in the figures the maximum count and minimum count drift in the middle of the protocol but eventually they all converge to $d_k$, where $k$ represents the maximum level.

## 5.3   Leader-driven, $O(\log^2 n)$-convergence-time exact size counting

In this section we show a $O(\log^2 n)$ time, high-probability protocol for a problem that is natural for agents with non-constant memory: exact population size counting. The probability of error can be reduced to 0 with standard techniques; see Corollary 5.9. This problem has been studied in the context of open protocols, in both the exact [18, 27] and approximate [18, 26] settings, where it is known that open protocols can approximate $n$ within multiplicative factor 2, by computing either $\lfloor \log n \rfloor$ or $\lceil \log n \rceil$, using $O(\log n \log \log n)$ states, and $O(\log^2 n)$ time [18], and open protocols can compute the **exact** value of $n$, using $O(n \log n \log \log n)$ states, and $O(\log n)$ time [18]. Both protocols can be changed to probability-1, with a multiplicative factor increase of $O(\log n)$ states in case, i.e., $O(\log^2 n \log \log n)$ states for calculating $\lfloor \log n \rfloor$ or $\lceil \log n \rceil$, and $O(n \log^2 n \log \log n)$ states for exactly computing $n$. However, note that our results below are leader-driven, so direct comparison with the leaderless results of [18] is not appropriate.

▶ **Theorem 5.7.** *There is an $O(1)$-message leader-driven population protocol (Protocol 3) that, with probability $1 - O(1/n)$, exactly counts the population size $n$ (storing it in each agent's internal state), in $O(\log^2 n)$ time and using $O(n \log^2 n)$ states.*

**Proof sketch.** It uses the "fast averaging" technique that has been useful in other population protocols [4, 18, 27, 39, 41], in which each agent holds an integer and computes the transition $i, j \to \lfloor \frac{i+j}{2} \rfloor, \lceil \frac{i+j}{2} \rceil$. In the $O(1)$-message setting, of course, this will not work exactly as described.

Intuitively, the leader will distribute 1 unit of what we can imagine is a continuous mass into the population. Rational-valued averaging of this mass would result in each agent converging to $1/n$, from which $n$ can be computed. $O(1)$ messages cannot represent arbitrary rationals. Instead, we allow agents to communicate a few bits of their number at a time, while ensuring that before moving on, they agree on an interval containing the true average, which shrinks by half each round, synchronized by a leader-driven phase clock [8]).

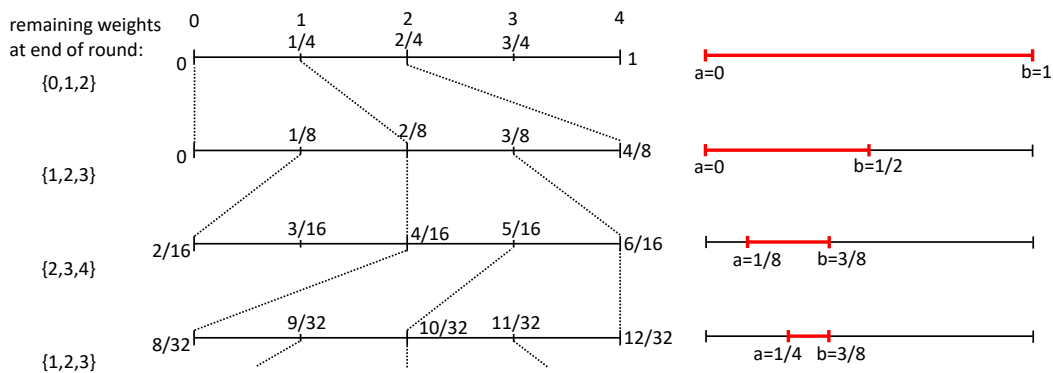

**Figure 3** Update rule for fast exact counting protocol. All agents start with a mass of 0 and weight $w = 0$, except the leader, who starts with mass $= 1$ and $w = 4$. They conduct averaging on weight $w$ for one round, at which point (with high probability) three consecutive weights remain. The figure shows how the remaining masses map to the next subinterval, with the weight $w$ updating to $2(w - w_{\min})$ where $w_{\min}$ is the minimum value of $w$ at the end of the *Averaging* phase. The right side shows the subintervals to scale. Each agent updates its internal state to represent the interval $[a, b]$. Once $[a, b]$ contains only a single number of the form $\frac{1}{n}$, the protocol terminates, and each agent knows the value $n$. The first $\log n$ rounds would always have 0 as the minimum remaining weights, but we allow other values to show concretely how the updating rule works.

Figure 3 shows the updating rule. Each agent's state represents an interval $[a, b] \subseteq [0, 1]$, where $b - a = 2^{-r}$ during round $r \in \mathbb{N}$ (initialized to $r = 0$). $a$ is a dyadic rational, initialized to $a = 0.0$, containing $r + 2$ bits after the binary point. One message field $W = \{0, 1, 2, 3, 4\}$ describes varying amounts of extra weight. The value $w \in W$ counts for $\frac{w}{4 \cdot 2^r}$ units of mass in round $r$. An agent is interpreted as having mass $= a + \frac{w}{4 \cdot 2^r} \in [a, b]$ (note representing $\frac{w}{4 \cdot 2^r}$ is what requires $r + 2$ bits after the binary point). The leader is initialized with $w = 4$ (and mass $= 0 + \frac{4}{4 \cdot 1} = 1$), and the followers are initialized with $w = 0$ (and mass $= 0$).

The full proof shows that, with high probability, every agent will always have the same value of $a$. This implies, via the averaging rule for weights, that mass is conserved and the sum of mass in the population is 1. Thus for all agents at all times, it holds that the true average $\frac{1}{n}$ stays within the interval $[a, b]$. Once the interval contains only a single integer reciprocal $\frac{1}{n}$, the protocol terminates with all agents knowing $n$. ◀

**Proof.** Protocol 3 shows pseudocode. The protocol works as follows. It uses the "fast averaging" technique that has been useful in other population protocols [4, 18, 27, 39, 41], in which each agent holds an integer and computes the transition $i, j \to \lfloor \frac{i+j}{2} \rfloor, \lceil \frac{i+j}{2} \rceil$. In the $O(1)$-message setting, of course, this will not work exactly as described.

Intuitively, the leader will distribute 1 unit of what we can imagine is a continuous mass into the population. Real-valued averaging of this mass would result in each agent converging to $1/n$, from which $n$ can be computed. Of course, we cannot store arbitrary-precision real numbers in states, though we can store rational approximations. But we cannot communicate arbitrary rationals using $O(1)$ messages. Instead, we allow agents to communicate a few bits of their number at a time, while ensuring that before moving on, they agree on an interval containing the true average, which shrinks by half each round (where synchronized rounds come from the leader-driven phase clock of [8]).

Each agent's state will represent an interval $[a, b] \subseteq [0, 1]$, where $b - a = 2^{-r}$ during round $r \in \mathbb{N}$ (initialized to $r = 0$). $a$ will be a dyadic rational, initialized to $a = 0.00$, containing $r + 2$ bits after the binary point. There is a message field $W = \{0, 1, 2, 3, 4\}$ describing varying amounts of extra weight. The value $w \in W$ counts for $\frac{w}{4 \cdot 2^r}$ units of mass in round $r$. An agent is interpreted as having mass $= a + \frac{w}{4 \cdot 2^r} \in [a, b]$ (note representing $\frac{w}{4 \cdot 2^r}$ is what requires $r + 2$ bits after the binary point). The leader is initialized with $w = 4$ (and mass $= 0.00 + \frac{4}{4 \cdot 1} = 1.00$), and the followers are initialized with $w = 0$ (and mass $= 0.00$).

We will prove that with high probability every agent will always have the same value of $a$. This will imply, via the averaging rule for weights, that mass is conserved and the sum of mass in the population is 1. Thus for all agents at all times, it holds that the true average $\frac{1}{n}$ stays within the interval $[a, b]$.

We first discuss the guarantees of the leader-driven phase clock from [8]. The agents will go through consecutive rounds, where each round contains an *Averaging* phase followed by an *Updating* phase. Choosing appropriate constant parameters, we can ensure with probability $1 - O(1/n)$ that for time $\Omega(n)$, all agents are synchronized within each phase for $\Theta(\log n)$ time, and no two agents are ever more than one phase apart (thus we can be sure that two agents in the same type of phase are also in the same round).

### *Averaging* phase

When two agents meet and both are in the *Averaging* phase, they each update just their weights via the standard averaging rule $i, j \to \lfloor \frac{i+j}{2} \rfloor, \lceil \frac{i+j}{2} \rceil$ for any $i, j \in W$. Note that assuming the invariant that every agent agrees on $a$, this rule preserves the sum of mass in the population.

Averaging takes time $\Theta(n)$ to converge in the worst case, where convergence happens when all agents agree on one of two consecutive integers $a$ and $a + 1$. (Thus further interactions are null.) However, Berenbrink, Friedetzky, Kaaser, and Kling [15] show that with probability $1 - O(1/n^2)$ it takes only $O(\log n)$ time to reach a configuration where all agents share **three** consecutive integers, two of which are $a$ and $a + 1$. The third could be either $a - 1$ or $a + 2$, depending on the true population-wide average; full convergence happens when all remaining $a - 1$'s encounter $a + 1$'s in the former case, and when all remaining $a + 2$'s all encounter $a$'s in the latter case. Thus, for appropriate constant parameters of the phase clock, with probability $1 - O(1/n)$, every *Averaging* phase lasts long enough such that at the end of each *Averaging* phase, we have $\max_w \leq \min_w + 2$.

### *Updating* phase

During the *Updating* phase, each agent spreads by epidemic the minimum weight $w_{\min}$ they have seen since the start of the *Updating* phase. By Lemma 4.4, we can guarantee with probability $1 - O(1/n)$ that every *Updating* phase lasts long enough for every agent to learn the minimum remaining message. (To ensure the $w_{\min}, w_{\min} + 1, w_{\min} + 2 \in W$ in case $w_{\min} > 3$ we take $\min(w_{\min}, 3)$).

At the end of the *Updating* phase, the agents update their lower bound to $a_{r+1} = a + \frac{w_{\min}}{4 \cdot 2^r}$. Because all agents agree on $w_{\min}$, they still agree on the value $a$ as desired. Assuming the weight in round $r$ was $w_r$, the weight updates to $w_{r+1} = 2(w_r - w_{\min})$. Because we have $\max_w \leq w_{\min} + 2$ after the successful *Averaging* phase, the set of weights at the start of every round will be in $\{0, 2, 4\}$ (see Figure 3). Notice that mass $= a_{r+1} + \frac{m_{r+1}}{4 \cdot 2^{r+1}} = a + \frac{m}{4 \cdot 2^r}$ is preserved during the update as desired.

Thus we have finished proving the invariant that all agents store the same interval $[a, b]$,

and the sum of mass is conserved, so $\frac{1}{n} \in [a, b]$. Finally, we can consider the time and space complexity, assuming that with probability $1 - O(1/n)$, this invariant holds and each round is $\Theta(\log n)$ time.

We first analyze the first round $r$ when the minimum weight $w_{\min} > 0$, so through this round we have $a = 0$ for all agents. Since the minimum weight $w_{\min} \geq 1$, the minimum mass is at least $0 + \frac{1}{4 \cdot 2^r} \leq \frac{1}{n}$, so $r \geq \log n - 2$. The maximum mass is at most $0 + \frac{4}{4 \cdot 2^r} \geq \frac{1}{n}$, so $r \leq \log n$. Thus we will have $w_{\min} > 0$ for the first time, increasing the lower bound $a$ after the *Updating* phase, at the end of round $r$, where $\log n \leq r \leq \log n + 2$. Corollary 5.8 argues how to use this fact to obtain a $O(\log n)$-state protocol for estimating $\log n$.

We next analyze the first round $r$ when the interval $[a, b]$ contains a unique reciprocal $\frac{1}{n}$. It is necessary and sufficient to have $\frac{1}{n+1} < a \leq b = a + 2^{-r} < \frac{1}{n-1}$. Thus it is necessary for $2^{-(r+1)} < \frac{1}{n-1} - \frac{1}{n+1} = \frac{2}{n^2-1}$, so $r > \log(n^2 - 1) - 1$. In the other direction, if round $r - 1$ did not uniquely determine $n$, then $2^{-(r-1)} \geq \frac{1}{n} - \frac{1}{n+1} = \frac{1}{n(n+1)}$, so $r \leq \log(n(n+1)) + 1$. Thus the protocol will terminate at the start of round $r = 2 \log n + O(1)$, and will take $O(\log^2 n)$ time.

Next, we analyze the space complexity. Naively storing $a$ (with $r + 2$ bits after the binary point) would use $2^{2 \log n + O(1)} = O(n^2)$ states. However, by the arguments given above, we have $a = 0$ until round $r_1 \approx \log n$, so simply store the counter $r_1$ to denote how many leading zeros $a$ has. The protocol will terminate at round $r_2 \approx 2 \log n$, so we can store $a$ with $\log n + \log \log n + O(1)$ bits. Including the counter $r$ and all constant space overhead gives a space bound of $\log n + 2 \log \log n + O(1)$ bits, so the total number of states is $O(n \log^2 n)$. ◂

Terminating Protocol 3 early gives a more space efficient protocol for estimating $\log n$:

▶ **Corollary 5.8.** *A leader-driven, $O(1)$-message protocol, with probability $1 - O(1/n)$, computes $r \in \{\lfloor \log n \rfloor, \lceil \log n \rceil\}$, in $O(\log^2 n)$ time using $O(\log n)$ states.*

**Proof sketch.** We run Protocol 3 until the interval $[a, b]$ contains exactly one power of two $2^{-k}$, and then output $k$, unless it contains no powers of two, in which case we output arbitrarily either of the powers of 2 contained in the interval of the **previous** round. If $n = 2^k$, then $k = \log n$ exactly. Otherwise, since the interval contains no other power of 2, but it contains $1/n$, then $k \in \{\lfloor \log n \rfloor, \lceil \log n \rceil\}$. ◂

**Proof.** We run Protocol 3 until the interval $[a, b]$ contains exactly one power of two $2^{-k}$, and then output $k$ (with one exception described below). If $n = 2^k$, then $k = \log n$ exactly. Otherwise, since the interval contains no other power of 2, but it contains $1/n$, then $k \in \{\lfloor \log n \rfloor, \lceil \log n \rceil\}$.

The interval $[a, b]$ endpoints are eventually arbitrary dyadic rationals. However, as observed in the proof of Theorem 5.7, in the first $\log n - 1$ rounds, the interval is of the form $[0, 2^{-j}]$, storable using only $O(\log n)$ states, because the minimum weight $w_{\min} = 0$. In the next round, depending on $w_{\min}$, the interval becomes one of $I_1 = [2^{-(j+2)}, 2^{-(j+2)} + 2^{-(j+1)}]$ (if $w_{\min} = 1$), or $I_2 = [2^{-(j+1)}, 2^{-j}]$ (if $w_{\min} = 2$). If $I_1$, then $2^{-(j+2)}$ is the only power of two in the interval.

If $I_2$, then one more round must pass. Note $I_2$ has only two powers of 2, the endpoints, and the interval will shrink to $I'_1$ in the next round, containing one of the endpoints (if $w = 0$ or 2), or neither (if $w = 1$). If $I'_1$ contains neither, then we know $n$ is not a power of 2, so we output $2^{-(j+1)}$ or $2^{-j}$ arbitrarily, since $j = \lfloor \log n \rfloor$ and $j + 1 = \lceil \log n \rceil$.

Only $O(1)$ extra states are needed to advance one more round, so $O(\log n)$ total states suffice. ◂

▌ **Protocol 3** *ExactCounting* (Agent $v$ seeing message $m$) is leader-driven with message fields leader $\in \{L, F\}$, $w, w_{\min} \in \{0, 1, 2, 3, 4\}$, and internal field $a \in \mathbb{Q}$. Subroutine LeaderDrivenPhaseClock gives internal field round $r \in \mathbb{N}$, message field phase $\in$ {*Averaging, Updating*}, and uses $O(1)$ message overhead to communicate the current phase.

---

**1 initial state of agent $v$:** $w \leftarrow 0$ if leader $= F$, $w \leftarrow 4$ if leader $= L$
  // w is a weight with a shrinking $2^{-2-r}$ units of mass
**2** $a \leftarrow 0.00$, $r \leftarrow 0$, phase $\leftarrow$ *Averaging* ;    // each agent has mass $= a + w/2^{2+r}$
**3** execute LeaderDrivenPhaseClock ;    // round r has phases *Averaging* then *Updating*
**4 if** $v.\text{phase} = m.\text{phase} = Averaging$ **then**
**5**     **if** $v$ is initiator **then**
**6**         $v.w \leftarrow \lceil \frac{v.w + m.w}{2} \rceil$ ;         // average both weights
**7**     **else**
**8**         $v.w \leftarrow \lfloor \frac{v.w + m.w}{2} \rfloor$
**9 if** *Averaging phase just ended* **then**
**10**    $v.w_{\min} \leftarrow \min(v.w, 3)$
**11 if** $v.\text{phase} = m.\text{phase} = Updating$ **then**
**12**    $v.w_{\min} \leftarrow \min(v.w_{\min}, m.w_{\min})$ ;    // learn the minimum weight in the population
**13 if** *Updating phase just ended* **then**
**14**    $v.a \leftarrow v.a + \frac{w_{\min}}{4 \cdot 2^r}$ ;     // update the mass interval lower bound
**15**    $v.w \leftarrow 2(v.w - v.w_{\min})$ ;   // update the weight to preserve mass for new round
**16**    **if** $[a, a + 2^{-r}]$ *contains a unique* $\frac{1}{n}$ **then**
**17**        terminate with population size $n$

---

By the standard technique of running in parallel with a slower deterministic counting protocol, we can convert Protocol 3 to have probability 0 of error while retaining fast convergence time.

▶ **Corollary 5.9.** *There are $O(1)$-message, leader-driven population protocols that, with probability 1, respectively count the exact population size $n$ and estimate it by computing $\lfloor \log n \rfloor$ or $\lceil \log n \rceil$, both with expected $O(\log^2 n)$ convergence time and $O(n \log^2 n)$ stabilization time. With probability $1 - O(1/n)$, they use $O(n^4 \log^4 n)$ and $O(\log^2 n)$ states, respectively.*

**Proof.** First consider the case of exact size counting. We can compose Protocol 3 with a slow stable counting algorithm. As a backup, we could use the deterministic broadcast mechanism sketched in Corollary 3.6, which stably counts the population with $O(n^3 \log^2 n)$ states in expected $O(n \log^2 n)$ time. Together with the $O(n \log^2 n)$ states of Protocol 3, this is $O(n^4 \log^4 n)$ states.

Because our protocol is leader-driven, we can use standard tricks to have the leader set a timer (see [8]) for when to tell all agents via epidemic to change their output to the deterministic backup. With probability $1 - O(1/n)$, the fast Protocol 3 will correctly compute $n$, and the timer will not go off until the backup has also stabilized. The probability $O(1/n)$ for errors add an expected $O((n \log^2 n)/n) = O(\log^2 n)$ convergence time to wait for the slow backup. Note the stabilization time is $\Omega(n \log^2 n)$ because until the slow backup has stabilized, there is a chance of switching to the backup before it is correct.

The case of size estimation is similar, although unlike the case of exact counting, in this paper we do not have a $O(1)$-message protocol that directly computes $\log n$ with probability 1. However, there is a simple open $O(\log n)$-state protocol that computes $\lfloor \log n \rfloor$: All agents start in state $\ell_1$, and for each $i$ and $j < i$ we have the transitions $\ell_i, \ell_i \to \ell_{i+1}, f_{i+1}$ and $f_i, f_j \to f_i, f_i$. This takes expected time $O(n)$ to elect a leader $\ell_{\lfloor \log n \rfloor}$ by the first type of transition and expected time $O(\log n)$ to propagate the value $\lfloor \log n \rfloor$ to all agents by epidemic.

Theorem 3.3 shows how any open, $s(n)$-state protocol can be simulated by a leader-driven, $O(1)$-message protocol with $O(n \log s(n))$ expected slowdown. This implies a leader-driven, probability-1 protocol for calculating $\lfloor \log n \rfloor$ with expected time $O(n^2 \log \log n)$. By combining this with the fast, error prone protocol described in Corollary 5.8, and setting the phase clock parameters of Protocol 3 to ensure probability of error at most $1/n^2$, the contribution to the expected time of the slow, probability-1 protocol is negligible, and the whole protocol runs in expected time $O(\log^2 n)$ time as in Corollary 5.8. It contributes $O(\log n)$ state complexity, so the total number of states is $O(\log^2 n)$. ◀

The next corollary shows that Protocol 3 can be made leaderless by composing with the leader election protocol derived from junta election (Protocol 2).

▶ **Corollary 5.10.** *There is a leaderless, $O(1)$-message population protocol that exactly counts the population size $n$ in $O(\log^2 n)$ time and $O(n \operatorname{polylog} n)$ states, succeeding with probability $1 - O(1/n)$. There is also a leaderless, $O(1)$-message population protocol that computes $\lfloor \log n \rfloor$ or $\lceil \log n \rceil$ in $O(\log^2 n)$ time and $O(\operatorname{polylog} n)$ states, succeeding with probability $1 - O(1/n)$.*

**Proof.** Both protocols work similarly. We can use *JuntaElection* (Protocol 2) to get a leader election protocol as in [32]. This will use $O(\log^2 n)$ state overhead and take $O(\log^2 n)$ parallel time. With probability $1 - O(1/n)$, all agents in Protocol 2 will restart when they enter the last level together with the same constant-factor estimate of $\log n$. They can use this estimate to set a timer to wait for the leader election to converge after $O(\log^2 n)$ time. Then we can start the downstream Protocol 3 to count the population, either exactly as in Theorem 5.7 or approximately as in Corollary 5.8. ◀

## 5.4 Leader-driven, $O(\log^2 n)$-time predicate computation

We can use techniques from Theorem 5.7 to show how to compute, using a leader and with high probability, any predicate on a constant alphabet $\Sigma$, up to the space bounds allowed by the agents. We assume that there is one leader agent, and that every other agent has a state from a fixed alphabet $\Sigma$. Exactly the semilinear predicates are computable with probability 1 by $O(1)$-state open protocols [7] (with $> \log n$ states, more predicates are possible [21]).

▶ **Corollary 5.11.** *Let $d \in \mathbb{N}^+$ and let $\Sigma$ be a d-symbol input alphabet. Then there is an $O(1)$ message leader-driven population protocol that, with probability $1 - O(1/n)$, exactly counts the input vector $\vec{i} \in \mathbb{N}^d$ (storing it in each agent's internal state), in $O(d \log^2 n)$ time and using $O(n^d \log^2 n)$ states.*

**Proof sketch.** Agents first run the Protocol 3 of Theorem 5.7 to store locally the value $n$. Agents then use a similar strategy to Protocol 3 to count how many agents have input $x$ for each symbol $x \in \Sigma$. Having now stored the entire initial population's input in their internal state, they can simply compute any computable predicate $\phi$ locally. ◀

**Proof.** First, agents run the Protocol 3 of Theorem 5.7 to store locally the value $n$. This protocol is terminating (i.e., agents signal when they are done and with high probability, no agent signals before all agents have converged), so it can be straightforwardly composed with the subsequently described protocol. Note that the state bound was $O(n \log^2 n)$, but we can store $n$ in a separate field that will only contribute $O(n)$ additional state overhead.

Next, we iterate over each element $x \in \Sigma$, counting the number of elements with symbol $x$ in the population, using the same transitions as Protocol 3, except now each agent storing $x$ starts with weight $w = 4$ and other agents start with $w = 0$. The same argument as the proof of Theorem 5.7, only now the total mass is $|x|$, and the agents will wait until the interval $[a, b]$ contains only one number $\frac{k}{n}$ for $k \in \mathbb{N}$. (Note this will only require an interval of length $O(\frac{1}{n})$ and thus take $\log n + O(1)$ rounds). After each of these sub-protocols terminates, the agents store the number $k = |x|$ in their internal state. Finally, note that the last element does not need to be counted, as it can recovered as the difference between $n$ and the counts of the other inputs.

It follows that this protocol will take $O(d \log^2 n)$ time and use $O(n^d \log^2 n)$ states. Alternately, we could run all these steps in parallel, which would reduce the time by a constant factor $d$, but increase the amount of messages used (to have $d$ independent copies of the weight and min weight fields $w, i$ for Protocol 3). The error probability follows from that of Theorem 5.7, taking a union bound over each of the $d+1$ instances of Protocol 3. ◀

If an agent can locally store the entire initial configuration, it can compute any predicate computable by the transition function $\delta$. Formally, we required that $\delta$ be computable by a Turing Machine with $O(\log s(n))$ bits of memory, to make our model comparable with [21]. Thus we can compute all predicates computable by $O(\log n)$ bit space-bounded Turing Machines via Corollary 5.11.

## 6 Computability with one-bit messages

We will show that with one-bit messages, it is possible to simulate a synchronous system that provides a one-bit broadcast channel. This in turn will be used to simulate more complex systems. The price is that we sacrifice stabilization for convergence, and rely on unbounded counters to ensure convergence in the limit with probability 1.

Let us begin by defining the simulated system. A **synchronous broadcast system** consists of $n$ synchronous agents that carry out a sequence of **rounds**. In a broadcast round, each agent generates a one-bit outgoing message. These outgoing messages are combined using the OR function to produce the outcome for this round.

Broadcast operations can be used to detect conditions such as the presence of a leader, or ordinary message transmission if a unique agent is allowed to broadcast in a particular round. However, because broadcast operations are symmetric, they cannot be used for symmetry breaking. For the purpose of electing a leader, we assume that agents have the ability to flip coins; once we have a leader, further agents may be recruited for particular roles using an auxiliary protocol that allows the leader to select a single agent from the population in some round. The broadcast and selection protocols are mutually exclusive: either all agents participate in a broadcast in some round or all agents participate in selection. This is possible by showing that all agents eventually agree on the round number forever with probability 1.

Simulating this model in a population protocol requires (a) enforcing synchrony across agents, so that each agent updates its state consistently with the round structure; (b) implementing the broadcast channel that computes the OR of the agents' outputs; and (c) implementing the selection protocol. We show how to do this in the following section.

## 6.1 Implementing the core primitives

Broadcasts are implemented by epidemics that propagate 1 messages, separated by barrier phases in which all agents display 0. Selection is implemented by having the leader display a 1 to the first agent it meets. Both protocols depend on the number of steps at each agent being approximately synchronized with high probability; after $t(n/2)$ steps, all agents' step counts should be within the range $t \pm O(\sqrt{t \log n})$ with high probability (see Lemma 4.1). The time to carry out a broadcast is also $O(\log n)$ with high probability (see Lemma 4.5). By increasing the length of each round over time, the total probability across all rounds of an error occurring in either the broadcast or selection protocol due to out-of-sync agents or slow broadcasts converges to a finite value. Applying the Borel-Cantelli lemma then shows that there is a round after which no further failures occur with probability 1.

### 6.1.1 Details

Observe that the probability that a particular agent $i$ participates in an interaction is exactly $2/n$, and that the events that $i$ participates in distinct interactions are independent. If we let $X_i^t$ be the indicator variable that agent $i$ participates in the $t$-th interaction, then $S_i^t = \sum_{j=1}^{t} X_i^t$ is a sum of independent Bernoulli random variables, and obeys the Chernoff bound $\Pr\left[|S_i^t - \mu| > \mu\delta\right] < 2e^{-\mu\delta^2/3}$, where $\mu = \mathrm{E}\left[S_i^t\right] = 2t/n$ and $0 \leq \delta \leq 1$.

The execution of each agent is organized as a sequence of rounds, where each round $r$ for $r = 1, 2, \ldots$ consists of exactly $5r^2$ steps. The first $2r^2$ steps will be a **barrier phase** during which the agent displays message 0 and updates its state during an interaction only by incrementing its step counter. The remaining $3r^2$ steps will be an **interaction phase** in which the agents may execute one of two protocols. In a **broadcast phase**, each agent will propagate an epidemic represented by message 1, recording if it observed such an epidemic and possibly initiating the epidemic itself if instructed to do so by the protocol. In a **selection phase**, a leader agent displays 1 for its first encounter, and the agent interacting with the leader receives a special mark. The choice of broadcast/selection phase is determined by the controlling protocol and is the same for all agents. As in a barrier phase, an agent in an interaction phase continues to update its step counter with each interaction.

The controlling protocol updates the state of the agent at the end of each round. Each agent $v$ has a state $v.\mathsf{state}$ that is one of broadcasting (agent is initiating a broadcast of value 1) receiving (agent is waiting to detect a 1), received (agent has detected a 1), selecting (agent is attempting to select another agent), candidate (agent is a candidate for selection), selected (agent has been selected), or idle (agent has selected another agent and is now waiting for the end of the round). We assume that the controlling protocol assigns consistent values to the agents in each phase: if one or more agents start in state broadcasting, the rest should start in state receiving; while if some agent starts in state selecting, the rest should start in state candidate. Pseudocode for the communication protocol is given in Algorithm 4 in 6.2.1, followed by a proof of its correctness in 6.2.2. The correctness of this protocol relies on carefully chosen phase interval lengths which allow us to simulate synchronous rounds wherein the above operations are executed in sequence.

## 6.2 Convergent Broadcast Algorithm

### 6.2.1 Algorithm Definition

**Algorithm 4** Convergent broadcast (Agent $v$ seeing message $m$)

```
1  v.tick ← v.tick + 1;
2  if v.tick < 2r² then
      // Barrier phase: do nothing
3  else if v.tick = 2r² and v.state = broadcasting then
      // End of barrier phase: start epidemic
4     v.m ← 1;
5  else if v.tick = 5r² then
      // End of interaction phase
6     Update v.state according to controlling protocol;
7     r ← r + 1;
8     v.tick ← 0;
9  else if v.tick > 2r² and v.state = receiving and m = 1 then
      // Receive and propagate epidemic
10    v.state ← received;
11    v.m ← 1;
12 else if v.tick = 3r² and v.state = selecting then
      // Attempt to select
13    v.m ← 1;
14 else if v.tick > 3r² and v.state = selecting and m = 0 then
      // Selected a candidate
15    v.m ← 0;
16    v.state ← idle;
17 else if v.tick > 2r² and v.state = candidate and m = 1 then
      // We are the selected candidate
18    v.state ← selected;
```

### 6.2.2 Proof of Correctness

Define $s_r = \sum_{j=1}^{r-1} 5r^2$; this is the total length of all rounds up to but not including $r$. Observe that $s_r = \Theta(r^3)$. Consider the midpoint $a_r = s_r + r^2$ of the barrier phase of round $r$. Let $A_{ir}$ be the event $|S_i^t - a_r| > r^2 - 1$, where $t = (n/2)a_r$ so that $\mathrm{E}\left[S_i^t\right] = a_r$. Then the Chernoff bound gives $\Pr\left[A_{ir} < 2e^{-a_r((r^2-1)/a_r)^2/3}\right] = e^{-\Theta(r)}$. Similarly define $b_r = s_r + 3r^2$ and $c_r = s_r + 4r^2$ as the steps 1/3 and 2/3 of the way through the interaction phase of round $r$, and define $B_{ir}$ as the event $|S_i^t - b_r| > r^2 - 1$ when $t = (n/2)b_r$ and $C_{ir}$ as the event $|S_i^t - c_r| > r^2 - 1$ when $t = (n/2)c_r$. Then we also have $\Pr\left[B_{ir}\right] = e^{-\Theta(r)}$ and $\Pr\left[C_{ir}\right] = e^{-\Theta(r)}$.

Finally, define $D_{ir}$ as the event that the schedule of interactions is such that an epidemic that has infected agent $i$ after $(n/2)b_r$ steps has not infected all agents after $(n/2)c_r$ steps. Note that this definition does not depend on whether an actual epidemic is in progress after $(n/2)b_r$ steps; instead, we consider a hypothetical epidemic starting at $i$ running on the same schedule. From Lemma 4.4, we have that for any two-way epidemic on $n$ processes, the expected value $\mathrm{E}\left[T_n\right]$ of the number of interactions to infect all agents is $O(n \log n)$ and $\Pr\left[T_n > (1+\delta)\mathrm{E}\left[T_n\right]\right] \leq 2.5 \ln(n) \cdot n^{-2\delta}$. For $D_{ir}$ to occur, we need $T_n > r^2$, giving $\delta = r^2/O(n \log n) - 1$ and thus $\Pr[D_{ir}] = e^{-\Omega(r^2/n \log n)} \ln n = e^{-\Omega(r^2/n \log n)}$.

Call a round $r$ **safe** if none of the events $A_{ir}, B_{ir}, C_{ir}$, or $D_{ir}$ occur for any $i \in \{1, \ldots, n\}$. These events are not even remotely independent, but the the union bound still applies, giving a probability that round $r$ is not safe of at most $3ne^{-\Omega(r)} + ne^{-\Omega(r^2/n \log n)} = e^{\Omega \log n - r} + e^{\Omega \log n - r^2/n \log n}$. The sum of of these bounds over all rounds converges to a finite value for any fixed $n$, so by the Borel-Cantelli lemma, with probability 1 all but finitely many rounds are safe.

The following lemmas demonstrate that the protocol does what it is supposed to, once we reach the suffix of the execution containing only safe rounds. We start by excluding false positive broadcasts.

▶ **Lemma 6.1.** *If rounds $r$ and $r+1$ are both safe, then no process observes a 1 in round $r$ unless some process initiates a broadcast or selection in round $r$.*

**Proof.** If rounds $r$ and $r+1$ are both safe, then the events $A_{ir}$ and $A_{i,r+1}$ do not occur for any $i$. In particular, this means that at time $t = (n/2)a_r$, all agents have an internal clock $S_i^t$ that is within the interval $a_r \pm r^2 - 1$, which lies within the barrier phase for round $r$. So at this time all agents display message 0, have completed the interaction phase for round $r-1$, and have not yet started the interaction phase for round $r$. A similar constraint holds at time $t' = (n/2)a_{r+1}$. It follows that any 1 observed by an agent during its round-$r$ interaction phase must result from some process setting its message to 1 either because it initiated an epidemic or selection during its own round-$r$ interaction phase, or because it is propagating an epidemic initiated by such a process. ◀

Similarly, a safe round has no false negative broadcasts:

▶ **Lemma 6.2.** *If round $r$ is safe and all agents start round $r$ in either a* broadcasting *or* receiving *state, then any epidemic initiated in round $r$ is observed by all agents.*

**Proof.** Because $B_{ir}$ does not occur for any $i$, after $(n/2)b_r$ steps, all agents are in their round-$r$ interaction phase, and because $C_{ir}$ does not occur for any $i$, all agents remain in their round-$r$ interaction phase until at least $(n/2)c_r$ steps. If some agent $i$ initiates an epidemic in round $r$, then $i.$state = broadcasting after $(n/2)b_r$ steps, and under the assumptions, of the lemma every other agent is either in the broadcasting, receiving, or received state. A simple induction shows that the set of infected agents in the real process throughout the $[(n/2)b_r, (n/2)c_r]$ interval is bounded below by the set of infected agents in the hypothetical epidemic considered in the definition of $D_{ir}$. This means that if $D_{ir}$ does not occur, both such sets contain all processes after $(n/2)c_r$ interactions. ◀

And a safe round allows selection. Selection is not necessarily uniform conditioned on safety, but each agent has an $\Omega(1/n)$ chance of being selected when $r$ is sufficiently large:

▶ **Lemma 6.3.** *If round $r$ is safe, exactly one agent $i$ starts round $r$ in a* selecting *state, and all other agents start round $r$ a* candidate *state, then exactly one agent finishes round $r$ in a* selected *state. For each agent $j$, the probability $p$ that its is chosen conditioned on the safety of round $r$ and the events of previous rounds is at least $\frac{1}{2n}$ for sufficiently large $r$.*

**Proof.** Use the non-occurrence of any $B_{ir}$ or $C_{ir}$ to argue that agent $i$ reaches tick $3r^2$ while all agents are in the interaction phase. Then the next interaction between $i$ and any $j$ causes $j$ to observe a 1 and switch to a selected state.

We would like to argue that the next interaction between $i$ and another agent $j$ chooses each $j$ with independent probability $1/n$. Unfortunately, we are conditioning on safety of round $r$. Let $A$ be the event that $i$ selects $j$ and $B$ the event that round $r$ is unsafe. Then $\Pr[A \mid \neg B] = \frac{\Pr[A \wedge \neg B]}{\Pr[\neg B]} \geq \Pr[A \wedge \neg B] = \Pr[A] - \Pr[A \wedge B] \geq \Pr[A] - \Pr[B] = 1/(n-1) - \left(e^{-\Omega(r)} + e^{-\Omega(r^2/n \log n)}\right) \geq \frac{1}{2n}$ for sufficiently large $r$. ◀

## 6.3 Convergent computation of arbitrary symmetric functions

Early rounds produce incorrect results, so we need an error-recovery mechanism. We describe a basic protocol for electing a leader and having it gather inputs from the other agents. This in principle allows the leader to compute the output of an arbitrary symmetric function and broadcast it to the other agents. The protocol guarantees termination with probability 1 even in executions where some of the rounds exhibit errors in the underlying broadcast mechanism. By restarting the protocol when it terminates, we can guarantee that the protocol eventually runs without errors, thus converging to the correct output.

Each agent $v$ maintains a Boolean field $v$.leader that marks it as a leader (or candidate leader) and a field $v$.processed that marks whether it has reported its input $v$.input to the leader. All agents rotate through a repeating sequence of 7 rounds, where the round number for the purposes of the protocol is $r \mod 7$. These are organized as follows:

**Round 0** Any leader broadcasts 1. A non-leader that receives 0 sets its leader bit. This round allows recovery from states with no leaders.

**Round 1** Any leader broadcasts 1 with probability 1/2. A leader that does not broadcast but receives a 1 clears its leader bit.

**Round 2** Any agent that cleared its leader bit in the previous round broadcasts 1. This causes any remaining leaders that receive a 1 to restart the information-gathering protocol and causes any non-leaders that receive a 1 to clear their processed bits. Broadcasting a 1 in this round is also used by the leader to restart the protocol after completion.

**Round 3** Any agent $v$ with $v$.processed $= 1$ broadcasts 1. This is used by the leader and other agents to detect unprocessed inputs.

**Round 4** If a leader received a 1 in the previous round, and there is no transmission in progress from a non-leader agent, the leader executes a selection operation. The selected agent sets its processed bit and transmits its input if its processed bit is not already set. If the processed bit is set, the agent transmits nothing in the following two rounds.

**Rounds 5 and 6** These are used to transmit either (a) one bit of a selected agent's input, or (b) one bit of the protocol output. In either case the bit is encoded as two bits using the convention $01 = 0$, $10 = 1$, $00 =$ stop. Note that the absence of a broadcast in both rounds is interpreted as stop, which both allows a selected agent to signal it has already been processed and guarantees eventual termination after an agent finishes transmitting its input even if some of the broadcasts are garbled.
It is possible for two agents to be transmitting simultaneously (this can occur if there are multiple surviving leaders). This requires that agents be prepared to handle receiving 11. The simplest way to handle 11 may be to have agents just interpret it as a fixed value: $11 = 1$. Alternatively, we could implement an optimization where any agent that observes 11 triggers a restart of the protocol by broadcasting a 1 in the next Round 2.

The protocol terminates when the leader has collected all inputs (detected by the absence of a signal in Round 3) and transmits the computed output to all agents (using Rounds 5 and 6 over however many iterations are needed). We assume that the computed output has finite length for any combination of inputs. After transmitting the output, the leader broadcasts a 1 in Round 2 to restart the information-gathering component of the protocol.

▶ **Lemma 6.4.** *In any execution with finite errors in the underlying broadcast protocol, with probability* 1*, the above protocol converges to a single leader and then restarts infinitely often.*

Lemma 6.4 is proven correct by demonstrating that, as defined, this protocol elects exactly one leader by Round 2 which correctly processes all non-leader agents with probability 1 by the end of Round 6. The full proof can be found in 6.4

## 6.4 Proof of Lemma 6.4

**Proof.** Consider a sequence of iterations in which no errors occur.

If there are no leaders initially, the first execution of Round 0 sets the leader bit in all agents. The only way that an agent can lose its leader bit is if it sees another leader broadcast a 1 in Round 1. But this always leaves at least one leader. If there is more than one leader, half the remaining leaders on average will drop out in each execution of Round 1. This guarantees that there will eventually be exactly one leader with probability 1.

If there is a leader, the leader believes that there is no transmission in progress, and at least one agent $v$ with $v.\mathsf{processed} = \mathsf{false}$, then $v$ is selected with probability at least $\frac{1}{2n}$ for sufficiently large $r$ in Round 3 (Lemma 6.3). This causes some agent to be selected in Round 3 eventually with probability 1, reducing the number of unprocessed agents by one.

If there is a transmission in progress, each transmitting agent sends finitely many bits before stopping. Once all transmitting agents have stopped, any agent waiting for a transmission to finish will observe 00 in Rounds 5 and 6.

It follows that starting from an arbitrary initial configuration, with probability 1 the protocol reaches a configuration with exactly one leader, the leader finishes waiting for any outstanding transmissions, and the leader then selects an unprocessed agent and collects its input until no unprocessed agents are left. After the leader transmits its computed (though possibly incorrect) output, the protocol restarts. ◀

Once the protocol restarts with a single leader, any subsequent error-free execution produces correct output. This follows from the fact that the leader collects the input from every agent exactly once. Since each agent records as its output the last output broadcast by the leader, this causes all agents to converge to holding the correct output with probability 1.

Because the leader has unbounded states, it can simulate an arbitrary Turing machine. This allows the output to converge to the value of any computable symmetric function. The restriction to symmetric functions follows from uniformity of the agents in the initial configuration, but can be overcome, assuming inputs include indexes. We have thus shown:

▶ **Theorem 6.5.** *For any computable symmetric function $f$, there is a population protocol using* 1*-bit messages and unbounded internal states that starts in an initial configuration where each agent $i$ is distinguished only by its input $x_i$, that converges to having each agent holding output $f(x_1, \ldots, x_n)$.*

Our construction exploits the unbounded state at each agent to allow the leader to simulate the entire computation. While the probability-1 convergence property requires unbounded state in the limit (otherwise there is a nonzero probability that any round fails), it may be desirable to put off expanding the state as long as possible. In 6.5, we argue that with some small tweaks, the construction can be adapted to distribute the contents of a Turing machine tape of $s$ bits across all agents of the population as in [21], reducing the storage overhead at each agent for the Turing machine computation to $O(s/n + \log s)$ bits.

## 6.5 Simulating a Turing machine

**Algorithm 5** TM simulation: leader

```
   // initial head position
 1 h ← 0;
   // initial state
 2 q ← q_0;
   // count of agents
 3 n ← 1;
   // indicates if n is correct
 4 counted ← false;
 5 Restart computation by sending 1 in Round 2;
   // Initialize tape
 6 while counted = false do
 7     if at least one agent reported processed = false in Round 3 then
 8         Select an agent in Round 4;
 9         Transmit Recruit(n + 1);
10         if some agent responds then
11             n ← n + 1;
12     else
           // No agents remaining!
13         Transmit PopulationSize(n);
14         counted ← true;
           // Initialize runtime bound
15         s ← f(n);

16 for each Turing machine step do
17     Read cell at current head position h by transmitting Read(h);
18     if some agent responds with c then
19         (q', c', d) ← δ(q, c);
20         Write c' to cell h by transmitting Write(h, c');
21         q ← q';
22         h ← h + d;
23         s ← s − 1;
24         if q is a halting state then
25             Transmit result of computation;
26             Jump to start of algorithm;
27         else if s = 0 then
               // Runtime bound exceeded
28             Jump to start of algorithm;
29     else
           // No agent holds h ⇒ initialization failed
30         Jump to start of algorithm;
```

**Algorithm 6** TM simulation: follower

```
   // initialization at restart
   // I am unallocated
 1 processed ← false;
   // j indexes which cells I hold
```
2 $j \leftarrow \bot$;
```
   // n is initially unknown
```
   // we use the convention that $x \bmod \infty = x$
   // and $x/\infty = 0$
3 $n \leftarrow \infty$;
   // $T[i]$ holds cell $(n-1)i + c$
   // Unset locations default to blank
4 Clear list $T[]$;
5 Set $T[0]$ to my input;
6 **upon receiving** Read($i$)
7    **if** $i \bmod (n-1) = j$ **then**
8       Transmit $T[\lfloor i/n \rfloor]$;
9 **upon receiving** Write($i, c$)
10   **if** $i \bmod (n-1) = j$ **then**
11      $T[\lfloor i/n \rfloor] \leftarrow c$;
12 **upon receiving** Recruit($h$)
13   **if** *I have been selected* **then**
14     **if** $j = \bot$ **then**
15       $j = h$;
16       Transmit acknowledgment;
17   Clear selected status;
18 **upon receiving** PopulationSize($m$)
19   $n \leftarrow m$;
20 **upon receiving** *result of computation*
21   Record result as output;

In this section, we show how to adapt the construction of Section 6.3 to simulate a Turing machine directly. For the most part, we retain the round structure of the previous construction, but make some adjustments to how the leader interacts with the other agents.

As in the construction in Section 6.3, we elect a leader using Rounds 0 and 1, which resets the other agents by broadcasting in Round 2. Non-leader agents reset to an **unallocated** state in which they hold only their input while waiting to be recruited to hold tape cells; such agents will set processed to false until recruited to hold a tape cell. The leader agent holds the state of the finite-state controller and the index for the current head position and manages communication with the other agents through Round 5 and 6 broadcasts. Using the same self-delimiting encoding as before allows transmission of messages of arbitrary length, so long as all agents agree on which agent's turn it is to speak.

A very high level overview of the simulation is given in Algorithms 5 and 6. Algorithm 5 is written from the perspective of the leader, and assumes that we have already elected a unique leader and reset all the other agents. The function $\delta : Q \times \Sigma \to Q \times \Sigma \times \{-1, 0, +1\}$ is the transition function for the simulated Turing machine. Algorithm 6 is written from the perspective of a non-leader and describes how it responds to transmissions from the leader.

The simulation starts by organizing the agents into a Turing machine tape. This involves selecting agents one at a time and assigning them indices. Because an agent might be selected more than once, the expected number of rounds to find all agents scales as $O(n \log^2 n)$, where $O(n \log n)$ comes from the expected time to finish a coupon collector process and the extra $O(\log n)$ comes from the time to transmit indexes one bit at a time. We assume that counting $n$ is also enough for the leader to compute a bound $f(n)$ on the number of steps used by the Turing machine; this is needed to enforce restarts if the simulated machine does not terminate on its own.

For simplicity we assume that inputs can be placed in arbitrary order on the first $n-1$ cells of the tape (this will require special handling of any input on the leader, which we omit for simplicity of presentation). A complication is that inputs to the protocol might exceed the size of the constant tape alphabet. This does not affect the simulation directly, since no restriction on tape alphabet is assumed, but it may require adding a preamble to the Turing machine computation that unpacks large-alphabet inputs into the constant-size TM alphabet. We leave the details of this tedious and unenlightening preamble to the imagine of the reader.

The analysis of Algorithms 5 and 6 essentially follows the proof of Theorem 6.5. Once the simulation reaches the safe phase of the construction, it reaches a configuration with one leader after some finite time with probability 1. At this point the leader may already have an inaccurate estimate $\hat{n}$ of $n$, but whether the estimate is accurate or not, each iteration of the main loop will require at most $O(\log f(\hat{n}))$ rounds to finish, leading to a restart after at most $O(n \log^2 n + f(\hat{n}) \log f(\hat{n}))$ rounds on average. Each subsequent iteration will run the Turing machine to completion and produce the correct output. The space complexity at each agent, measured in bits, is bounded by $O(s/n + \log s)$ during non-faulty simulations, where $s$ is the largest tape cell index used. For faulty executions, we accept a small probability that a larger estimate of $\hat{n}$ at some leader agent may leader to larger space overhead. In either case the state complexity is dominated in the limit by the unbounded round and tick counters.

▶ **Theorem 6.6.** *Algorithms 5 and 6 use the synchronous broadcast primitive to simulate a Turing machine with known time complexity $f(n)$, converging to the correct output with probability* 1. *In any execution, the additional space required at each agent to simulate a Turing machine that uses $s$ tape cells is bounded after an initial prefix by $O(s/n + \log s)$ with probability* 1.

## 7  Open problems

**Probability-1 computation.** Many of our protocols have a positive probability of error. Common techniques for achieving zero error probability in $\omega(1)$-state protocols require $\omega(1)$ messages. Based on this, we conjecture that probability-1 leader election using $O(1)$ messages requires $\Omega(n)$ time to stabilize. This is known to hold for $O(1)$ states [29], though sublinear-time **convergence** is possible with $O(1)$ states [37].

**Time lower bounds.** A tool for time lower bounds (e.g., probability-1 leader election [1,29]) is a "density lemma" [1,25] showing that when the state complexity is $\leq \frac{1}{2} \log \log n$, all states appear in "large" count. This is false for $s(n) > \log \log n$, which is the key to the fastest space-optimal leader election protocols [16, 32, 33]. A density lemma applies to the **messages** of $O(1)$-message protocols, no matter the state complexity (derivable from [26, Lemma 4.2]). Does this imply that $O(1)$-message leader election requires linear stabilization time?

**Power of 1-bit messages with $O(1)$-states.** $O(1)$-state **open** protocols can stably compute exactly the semilinear predicates [7]. Can all semilinear predicates be stably computed with 1-bit messages? A related question is whether there is a direct simulation of $O(1)$-message protocols by 1-bit message protocols (similar to Theorem 3.3).

**Efficient predicate computation.** Corollary 5.11 can be used to efficiently compute any computable predicate $\phi : \mathbb{N}^d \to \{0, 1\}$ but requires storing the entire initial configuration locally in each agent ($\Theta(n^d)$ states). Corollary 3.6 can be used to compute any computable predicate storing unique IDs in each agent ($O(n)$ states), but it is slow since communication is routed through a leader. What predicates can be computed time- *and* space-efficiently?

**Acknowledgements.**

David Doty, Mahsa Eftekhari, and Eric Severson were supported by NSF grants 1619343, 1844976, 1900931. James Aspnes was supported in part by NSF grant CCF-1650696.